\documentclass[a4paper,12pt]{article}

\usepackage{amsmath,times,amssymb,latexsym,multicol,graphics}
\usepackage{ascmac,fancybox}
\usepackage{bbm}
\usepackage{wrapfig}
\usepackage{framed}
\usepackage{niceframe}
\usepackage{setspace}
\usepackage{comment}
\usepackage{authblk}
\usepackage{braket}
\usepackage{simplewick}
\usepackage{tikz}
\usepackage{color}\usepackage{here}

\newcommand{\be}{\begin{equation}}
\newcommand{\ee}{\end{equation}}

\newcommand{\e}{\epsilon}

\newcommand{\la}{\lambda}

\newcommand{\bpm}{\begin{pmatrix}}
\newcommand{\epm}{\end{pmatrix}}
\newcommand{\bbm}{\begin{bmatrix}}
\newcommand{\ebm}{\end{bmatrix}}

\usepackage{cite}

\usepackage{hyperref}

\begin{document}
\begin{titlepage}
\begin{flushright}
MPP-2020-040, YITP-20-42\\
{\today}
 \\
\end{flushright}

\begin{center}

\vspace{1cm}

\hspace{3mm}{\LARGE \bf A Field Theory Study of Entanglement Wedge Cross Section: Odd Entropy} \\[3pt] 

\vspace{1cm}

\renewcommand\thefootnote{\mbox{$\fnsymbol{footnote}$}}
Ali {Mollabashi}${}^{1}$, Kotaro {Tamaoka}${}^{2}$

\vspace{5mm}
${}^{1}${\small \sl Max-Planck-Institut for Physics}\\
{\small \sl Werner-Heisenberg-Institut 80805 Munich, Germany}\\
${}^{2}${\small \sl Center for Gravitational Physics, 
Yukawa Institute for Theoretical Physics (YITP), Kyoto University, } \\ 
{\small \sl Kitashirakawa Oiwakecho, Sakyo-ku, Kyoto 606-8502, Japan}\\

\vspace{5mm}
${}^{1}${\small{\,alim@mppmu.mpg.de}}\\
${}^{2}${\small{\,kotaro.tamaoka@yukawa.kyoto-u.ac.jp}
}
\end{center}

\vspace{5mm}

\noindent
\abstract
We study odd entanglement entropy (odd entropy in short), a candidate of measure for mixed states holographically dual to the entanglement wedge cross section, in two-dimensional free scalar field theories. Our study is restricted to Gaussian states of scale-invariant theories as well as their finite temperature generalizations, for which we show that the odd entropy is a well-defined measure for mixed states. Motivated from holographic results, the difference between odd and von Neumann entropy is also studied. In particular, we show that large amounts of quantum correlations ensure the odd entropy to be larger than von Neumann entropy, which is qualitatively consistent with the holographic CFT. In general cases, we also find that this difference is not even a monotonic function with respect to size of (and distance between) subsystems. 

\end{titlepage}
\setcounter{footnote}{0}
\renewcommand\thefootnote{\mbox{\arabic{footnote}}}
\tableofcontents
\flushbottom

\section{Introduction and Summary}
\subsection{Introduction}
One of the main interests in quantum (many-body) physics is the structure of entanglement in a given density matrix. Entanglement entropy, the unique measure of the quantum entanglement for pure states, is a useful instrument to tackle this problem. For general mixed states, however, we have no unique candidate to approach this problems for now. 

In light of the AdS/CFT correspondence\cite{Maldacena:1997re}, the entanglement entropy gives a sharp geometrical counterpart, the area of the minimal surfaces\cite{Ryu:2006bv}. This relation, called the Ryu-Takayanagi formula, gives us a profound connection between entanglement in CFTs and geometries in asymptotically AdS spacetimes. With the fact mentioned in the above paragraph in mind, we can expect that we would have some generalization of this formula. Such generalizations should elucidate the relation between the structure of correlations for mixed states and a subregion in bulk spacetime, namely the so-called entanglement wedge\cite{Wall:2012uf, Czech:2012bh, Headrick:2014cta}. 

Recently, a generalization of the minimal surfaces associated with the entanglement wedge has been introduced, dubbed the entanglement wedge cross section\cite{Takayanagi:2017knl, Nguyen:2017yqw}. 
A natural question is what is the boundary dual of the entanglement wedge cross section? After the original proposal on the entanglement of purification\cite{Takayanagi:2017knl, Nguyen:2017yqw}, we have also received many possible candidates: logarithmic negativity\cite{Kudler-Flam:2018qjo,Kusuki:2019zsp}, odd entanglement entropy (often called odd entropy in short)\cite{Tamaoka:2018ned}, reflected entropy\cite{Dutta:2019gen}, and $R$-correlation\cite{Umemoto:2019jlz,Levin:2019krg}. 
Since the Einstein gravity limit is quite universal, it would be possible that many different measures for mixed states are related holographically to the same geometrical object\footnote{To be precise, the ways to relate these quantities to the entanglement wedge cross section are different among each other. The case of odd entropy, the main target of this paper, will be reviewed in the next subsection. In the two-dimensional CFT case, this universality can be also seen explicitly from the conformal block at the large $c$ limit\cite{Hirai:2018jwy}. }. 

However, this is not true for more general quantum systems. Comparing the holographic results with each quantity in more general systems, we could learn different perspectives on how holographic CFTs should be characterized. In other words, these quantities can be useful to classify the many-body systems (including quantum gravity via holography) in various senses. 

In this paper, based on this expectation, we numerically study the odd entropy for free scalar field theories by using the harmonic chains in various setups (see section \ref{sec:harmonic}). Since very little is known about the odd entropy itself, this is indeed the first step towards the understanding of generic features of this quantity.  

\subsection{Review of odd entropy}
We review the odd (entanglement) entropy briefly. Let $\rho_{AB}$ ($AB\equiv A\cup B$) be a density matrix acting on bi-partite Hilbert space $\mathcal{H}_A\otimes\mathcal{H}_B$. First, we introduce the partial transposition $T_B$\cite{Peres:1996dw}, 
\be
\bra{k_A,\ell_B}\rho^{T_B}_{AB}\ket{m_A,n_B}=\bra{k_A,n_B}\rho_{AB}\ket{m_A,\ell_B}.
\ee
Since the transposition is not a completely positive map, the partial one can lead to negative eigenvalues. Then, the odd entropy is defined as
\be
S_o(A:B)\equiv\sum_{i}\mathrm{sign}(\la_i)(-|\la_i|\log|\la_i|),
\ee
where $\la_i$s are eigenvalues of $\rho_{AB}^{T_B}$. In particular, we have $S_o(A:B)=S(A)=S(B)$ if $\rho_{AB}$ is a pure state. Here, $S(A)$ denotes the entanglement entropy for subregion $A$. We are also interested in the difference between odd entropy and von Neumann entropy, $S_o(A:B)-S(AB)$. We expect this difference gives the area of the minimal cross section of the entanglement wedge, so-called entanglement wedge cross section\cite{Takayanagi:2017knl, Nguyen:2017yqw},
\be 
S_o(A:B)-S(AB)=E_W(A:B). \label{eq:oeew}
\ee 
In the two-dimensional holographic CFTs, one can explicitly confirm this relation by using the replica trick\cite{Tamaoka:2018ned,Kusuki:2019rbk,Kusuki:2019evw,Kudler-Flam:2020url}. Since it is related to an area of a certain surface, we should have $S_o(A:B)\geq S(AB)$ at the leading order of the large-$N$ limit. One may expect this inequality always holds, however, we can easily find counterexamples of the inequality from mixed states in two qubit systems, for example (see appendix \ref{app:twoq}). Therefore, it is worth to understand {\it when} this inequality holds in more generic setups. This is the question we address in the present paper.

\subsection{Summary of results}
\begin{figure}[t]
 \begin{center}
 \resizebox{120mm}{!}{\includegraphics{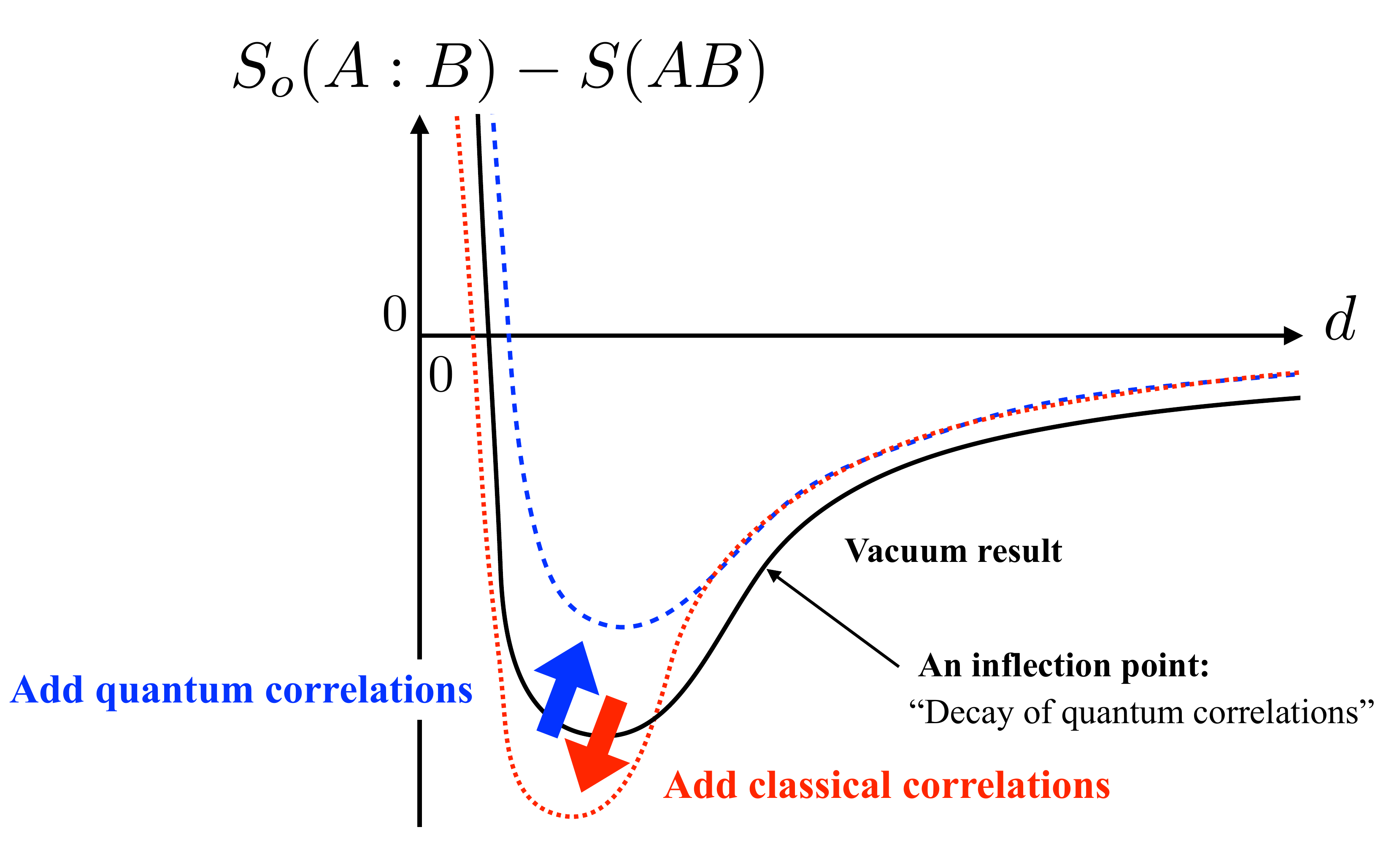}}
 \end{center}
 \caption{Qualitative behavior of the difference between $S_o(A:B)$ and $S(AB)$ in our setup. The horizontal axis corresponds to the distance between two subsystems $A$ and $B$. The minimum value increases/decreases as we ``add'' further quantum/classical correlations. We also find an inflection point which reflects the decay of quantum correlations. }
  \label{fig:summary}
\end{figure}

In this paper, we study the odd entropy for reduced density matrices of the vacuum in two dimensional free scalar fields (section \ref{sec:vac}).
We also consider thermal states (section \ref{subsec:thm}) and its non-relativistic generalization called Lifshitz theories (section \ref{subsec:lif}). The reason is as follows: we can regard thermal state/Lifshitz vacuum state as the one parameter generalization of the original vacuum state for the free scalar fields. In particular, these deformation increase \textit{classical}/\textit{quantum} correlations. We are interested in how odd entropy is different from the usual entropy via such deformations. Remarkably, odd entropy turns out to become larger than the von Neumann entropy as we increase {\it quantum} correlations (see Figure \ref{fig:summary}). On the other hand, it becomes smaller as we increase classical correlations. Note that, in holographic CFTs, we have large amount of quantum correlations\cite{Hayden:2011ag} (see also recent developments based on the entanglement wedge cross section\cite{Umemoto:2019jlz,Akers:2019gcv}). This observation qualitatively answers the previous question, why the difference between odd entropy and von Neumann entropy is always positive semi-definite for holographic CFTs. 

Moreover, we will further confirm numerically the following (unproved) properties (see section \ref{sec:ineq}):
\begin{itemize}
\item $S_o(A:B)\geq0$ (positive semi-definiteness) \vspace{2mm}
\item $S_o(A:B_1B_2)\geq S_o(A:B_1)$ (monotonicity) \vspace{2mm}
\item $S_o(A:B_1B_2)\leq S_o(A:B_1)+S(A:B_2)$ (polygamy relation) \vspace{2mm}
\item $S_o(A_1A_2:B_1B_2)\geq S_o(A_1:B_1)+S(A_2:B_2)$ (breaking of strong super additivity)
\end{itemize}
These observations ensure that the odd entropy is a well-defined measure for mixed states.  

The organization of the rest of this paper is as follows. In section \ref{sec:harmonic}, we first explain technical aspects of our setup in terms of discretized models, harmonic chains. In section \ref{sec:vac}, we study the odd entropy for reduced density matrix obtained from the vacuum. Then, we further study the same setup in thermal states (section \ref{subsec:thm}) and its Lifshitz-type generalizations (section \ref{subsec:lif}). In section \ref{sec:ineq}, we explore some inequalities (monotonicity, monogamy/polygamy relation and strong superadditivity) of odd entropy. In section \ref{sec:discussion}, we conclude with some discussion, especially comparison with the holographic results. In appendix \ref{app:inf}, we discuss inflection points in $S_o-S$ further based on two qubit model and thermal state with single interval. We also leave some plots in appendix \ref{app:MI} to compare the odd entropy with other measures. 

\section{Basic setup in harmonic chain}\label{sec:harmonic}
This is a technical section to explain what kind of models we will consider in this paper and how we can compute the odd entropy thereof. The reader who is not interested in the technical details may skip this section. 

We would like to study the odd entropy for free scalar field theories numerically. To this end, we consider the one-dimensional harmonic chain which is a series of the harmonic oscillators $(q_n,p_n)$. The Hamiltonian is given by,
\be
H=\sum_{n=1}^{N}\left[\dfrac{1}{2M}p^2_n+\dfrac{M\omega^2}{2}q^2_n+\dfrac{K}{2}(q_{n+1}-q_n)^2\right]. \label{eq:harmonicH}
\ee

\begin{figure}[t]
 \begin{center}
 Line with Dirichlet boundary condition\vspace{2mm}\\
 \resizebox{140mm}{!}{
 \includegraphics{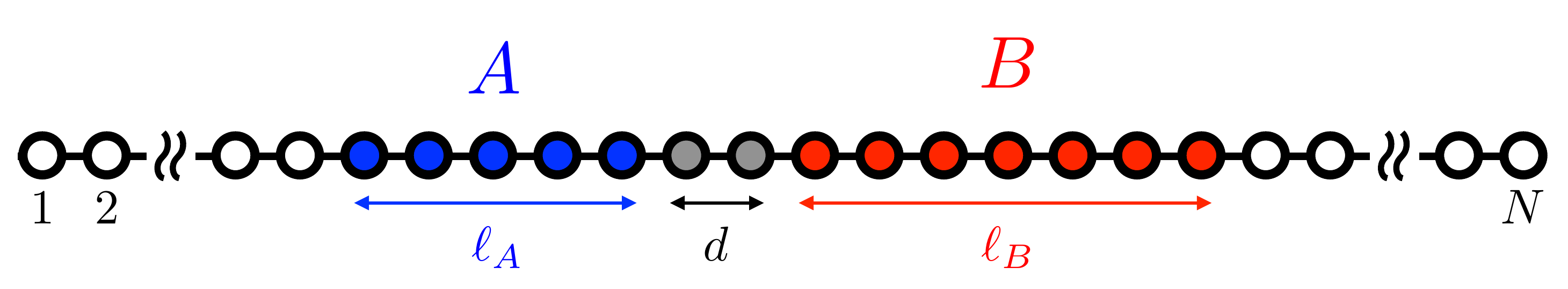}}
 \end{center}
 \caption{Our setup for $1$d lattice with $N$ sites. We mainly focus on Dirichlet boundary condition and set the lattice spacing $a=1$ for simplicity. We prepare the subsystem $AB\equiv A\cup B$ by tracing over its compliment. In what follows, $\ell_A$ and $\ell_B$ denote the total number of the site (namely, length) for each subsystem $A$ and $B$. We also use $d$ as distance between $A$ and $B$. In this figure, we have $\ell_A=5$, $\ell_B=6$, and $d=2$. Since now we are not interested in the boundary effect, we put $AB$ at the center and take $N$ large so that $\ell_A+d+\ell_B\ll N$. }\label{fig:setup}
\end{figure}

\begin{figure}[t]
 \begin{center}
  Circle with periodic boundary condition\vspace{2mm}\\
 \resizebox{70mm}{!}{\includegraphics{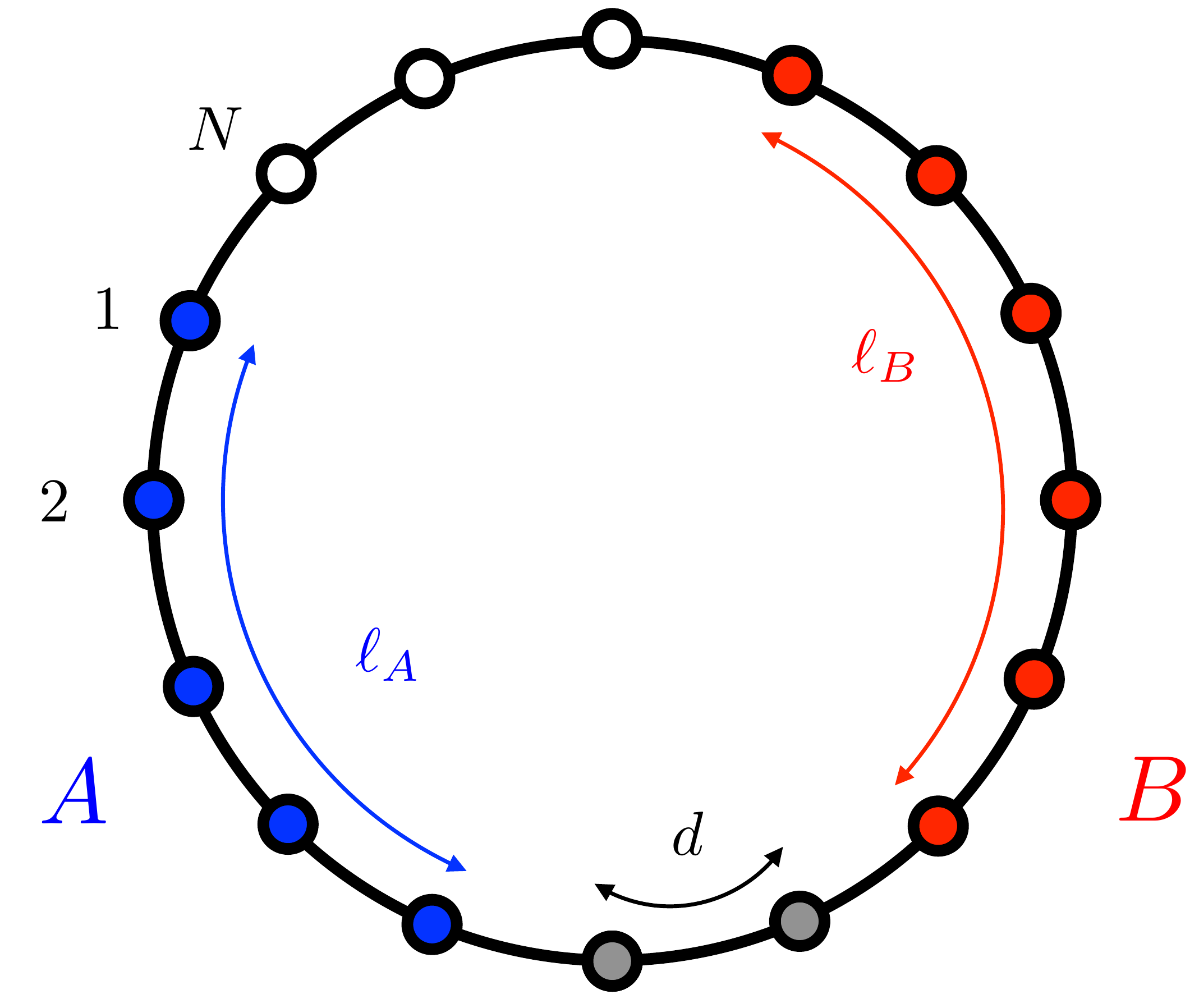}}
 \end{center}
 \caption{We also study the case with periodic boundary condition. In this case, as we will comment in the next subsection, we need to introduce non-zero mass term into the theory in order to avoid the zero-modes. }
  \label{fig:setup2}
\end{figure}

We will consider the Dirichlet boundary condition $q_1=q_N=p_1=p_N=0$ (Figure \ref{fig:setup}). and periodic boundary conditions $q_1=q_{N+1}, p_1=p_{N+1}$ (Figure \ref{fig:setup2}). These variables obey the canonical commutation relation $[q_n,p_m]=i\delta_{nm}$ and $(\textrm{others})=0$. After rescaling the variables,
\be
(q_n,p_n)\rightarrow((MK)^{1/4}q_n, (MK)^{-1/4}p_n),
\ee
we have only two parameters,
\be
H=\sum_{n=1}^{N}\left[\dfrac{1}{2a}p^2_n+\dfrac{a\omega^2}{2}q^2_n+\dfrac{1}{2a}(q_{n+1}-q_n)^2\right].
\ee
Here we introduced $a\equiv \sqrt{M/K}$. This Hamiltonian corresponds to the lattice free scalar field with the lattice spacing $a$ and mass $\omega$. The continuum limit can be realized as
\be
q_n\rightarrow \phi(x), p_n/a\rightarrow \pi(x)=\dot{\phi}(x), 
\ee
with $x=na, Na=L$ and $a\rightarrow0, N\rightarrow \infty$. 
In below, we set $M=K=a=1$ for simplicity. We have the following dispersion relation with respect to the Fourier mode $k$,
\be
\omega_k=\sqrt{\omega^2+4\left(\sin\frac{\pi k}{2N}\right)^2},\hspace{2mm} (k=1,\cdots,N-1), 
\ee
for Dirichlet boundary condition and 
\be
\omega_k=\sqrt{\omega^2+4\left(\sin\frac{\pi k}{N}\right)^2},\hspace{2mm} (k=0,\cdots,N-1).
\ee
for periodic boundary condition. Note that we have zero modes in the latter case. Therefore, we need to introduce small-mass parameter when we want to consider the CFT (massless) limit. As far as $\omega L=\omega N\ll1$ is satisfied, we can regard the numerical result as massless (conformal) limit.

We can compute the odd entropy for Gaussian states from the correlation functions as like the logarithmic negativity\cite{Calabrese:2012ew, Calabrese:2012nk}. In that, we can compute trace of the reduced density matrices from the eigenvalues of so-called covariance matrices. For a review, please refer to \cite{Casini:2009sr}. These matrices, we will spell out explicitly in below for each setup, are determined by correlation functions for $q_n$'s and $p_n$'s with respect to a given state of interest.

\noindent
\textbf{Vacuum state:} In this case, the correlation functions are given by
\begin{align}
\braket{0|q_mq_n|0}&=\dfrac{1}{2N}\sum^{N-1}_{k=0}\dfrac{1}{\omega_k}\cos\left(\dfrac{2\pi k(n-m)}{N}\right)\equiv\mathbb{Q}^{(p)}_{mn} \\
\braket{0|p_mp_n|0}&=\dfrac{1}{2N}\sum^{N-1}_{k=0}\omega_k\cos\left(\dfrac{2\pi k(n-m)}{N}\right)\equiv\mathbb{P}^{(p)}_{mn}
\end{align}
for periodic chain and
\begin{align}
\braket{0|q_mq_n|0}&=\dfrac{1}{N}\sum^{N-1}_{k=0}\dfrac{1}{\omega_k}\sin\left(\dfrac{\pi k m}{N}\right)\sin\left(\dfrac{\pi k n}{N}\right)\equiv\mathbb{Q}^{(D)}_{mn}, \label{eq:Dvacqq}\\
\braket{0|p_mp_n|0}&=\dfrac{1}{N}\sum^{N-1}_{k=0}\omega_k\sin\left(\dfrac{\pi k m}{N}\right)\sin\left(\dfrac{\pi k n}{N}\right)\equiv\mathbb{P}^{(D)}_{mn}, \label{eq:Dvacpp}
\end{align}
for Dirichlet chain. Then, the trace of $n$-th power of reduced density matrix is given by
\be
\mathrm{Tr}\rho^n_{AB}=\prod^{|AB|}_{i=1}\left[\left(\mu_i+\dfrac{1}{2}\right)^n-\left(\mu_i-\dfrac{1}{2}\right)^n\right]^{-1},
\ee
where each $\mu_i$ denoted the (square root of) eigenvalues for the matrix $\mathbb{Q}_{AB}\cdot\mathbb{P}_{AB}$,
\be
\mathrm{Spec}(\mathbb{Q}_{AB}\cdot\mathbb{P}_{AB})=\{\mu^2_1,\cdots,\mu^2_{|AB|}\}.
\ee
In the same way, one can compute the partially transposed one as
\be
\mathrm{Tr}(\rho^{T_B}_{AB})^n=\prod^{|AB|}_{i=1}\left[\left(\nu_i+\dfrac{1}{2}\right)^n-\left(\nu_i-\dfrac{1}{2}\right)^n\right]^{-1},
\ee
where we defined
\begin{align}
\mathrm{Spec}(\mathbb{Q}_{AB}\cdot\mathbb{P}^{T_B}_{AB})&=\{\nu^2_1,\cdots,\nu^2_{|AB|}\},\\
\mathbb{P}^{T_B}_{AB}&=\mathbb{R}_B\mathbb{P}_{AB}\mathbb{R}_B,\\
(\mathbb{R}_B)_{mn}&=\delta_{mn}(-1)^{\delta_{m\in B}}.
\end{align}
Therefore, the odd entropy is computed as
\be
S_o(A:B)=\sum^{|AB|}_{i=1}\left[\left(\nu_i+\dfrac{1}{2}\right)\log\left(\nu_i+\dfrac{1}{2}\right)-\mathrm{sign}\left(\nu_i-\dfrac{1}{2}\right)\left|\nu_i-\dfrac{1}{2}\right|\log\left|\nu_i-\dfrac{1}{2}\right|\right].
\ee
\noindent
\textbf{Thermal states:} 
In this case, we can simply replace the vacuum correlators to thermal expectation values:
\begin{align}
\mathbb{Q}_{mn}&=\dfrac{1}{L}\sum^{L-1}_{k=1}\dfrac{1}{\omega_k}\coth\left(\dfrac{\beta\omega_k}{2}\right)\sin\left(\dfrac{\pi k m}{L}\right)\sin\left(\dfrac{\pi k n}{L}\right), \label{eq:thmqq}\\
\mathbb{P}_{mn}&=\dfrac{1}{L}\sum^{L-1}_{k=1}\omega_k\coth\left(\dfrac{\beta\omega_k}{2}\right)\sin\left(\dfrac{\pi k n}{L}\right)\sin\left(\dfrac{\pi k m}{L}\right).\label{eq:thmpp}
\end{align}
Technically, all we need to do is just repeating the previous procedures by using the above \eqref{eq:thmqq} and \eqref{eq:thmpp} instead of \eqref{eq:Dvacqq} and \eqref{eq:Dvacpp}. 

\noindent
\textbf{Lifshitz vacuum states:}
The model is defined very similar to \eqref{eq:harmonicH} except that the interacting term between oscillators is replaced as\cite{MohammadiMozaffar:2017nri,He:2017wla}
\be
(q_{n+1}-q_n)^2 \rightarrow \left(\sum^z_{k=0}(-1)^{z+k}{}_zC_kq_{n-1+k}\right)^2, 
\ee
where now we have a longer range interaction proportional to $z$. Entanglement structure of these models has been further studied in \cite{MohammadiMozaffar:2017chk, MohammadiMozaffar:2018vmk}. The specific choice for such an interaction term is made to recover a Lifshitz invariant scalar theory in the continuum. The diagonalization of this model is again achieved by simply taking the Fourier transformation, this time with the following dispersion relation for Dirichlet boundary condition
\be
\omega_k=\sqrt{\omega^{2z}+4\left(\sin\dfrac{\pi k}{2N}\right)^{2z}}
\ee
and a similar one for the periodic chain. The correlation functions follow the same expressions with the aforementioned generalized dispersion relation. 

\section{Odd entropy for vacuum states}\label{sec:vac}

\subsection{Two adjacent interval}\label{subsec:vac_adj}
If the subregion $A$ and $B$ are adjacent each other ($d=0$ case in Figure \ref{fig:setup}), the CFT${}_2$ result is universally given by
\begin{align}
S_o(A:B)&=S(AB)+\dfrac{c}{6}\log\dfrac{\ell_A\ell_B}{\e(\ell_A+\ell_B)}+\textrm{const.}\,,\label{eq:adj}\\
S(AB)&=\dfrac{c}{3}\log\dfrac{(\ell_A+\ell_B)}{\e},
\end{align}
where $\ell_{A,B}$ are size of the subsystems $A,B$, $\e$ is UV cutoff, and $c$ is the central charge. In our case, we have $c=1$ and set $\e=1$. The constant term depends on the detail of theories which is not important now. This expression has been obtained from an alternative expression of the odd entropy, 
\be
S_o(A:B)=\lim_{n_o\rightarrow1}\dfrac{\mathrm{Tr}\rho^{T_B}_{AB}-1}{1-n_o}.
\ee
Here we took $n_o$ as an analytic continuation of {\it odd} integer. 
This expression is particularly useful when we compute the odd entropy in field theories. 

\begin{figure}[t]
 \begin{center}
 \resizebox{100mm}{!}{\includegraphics{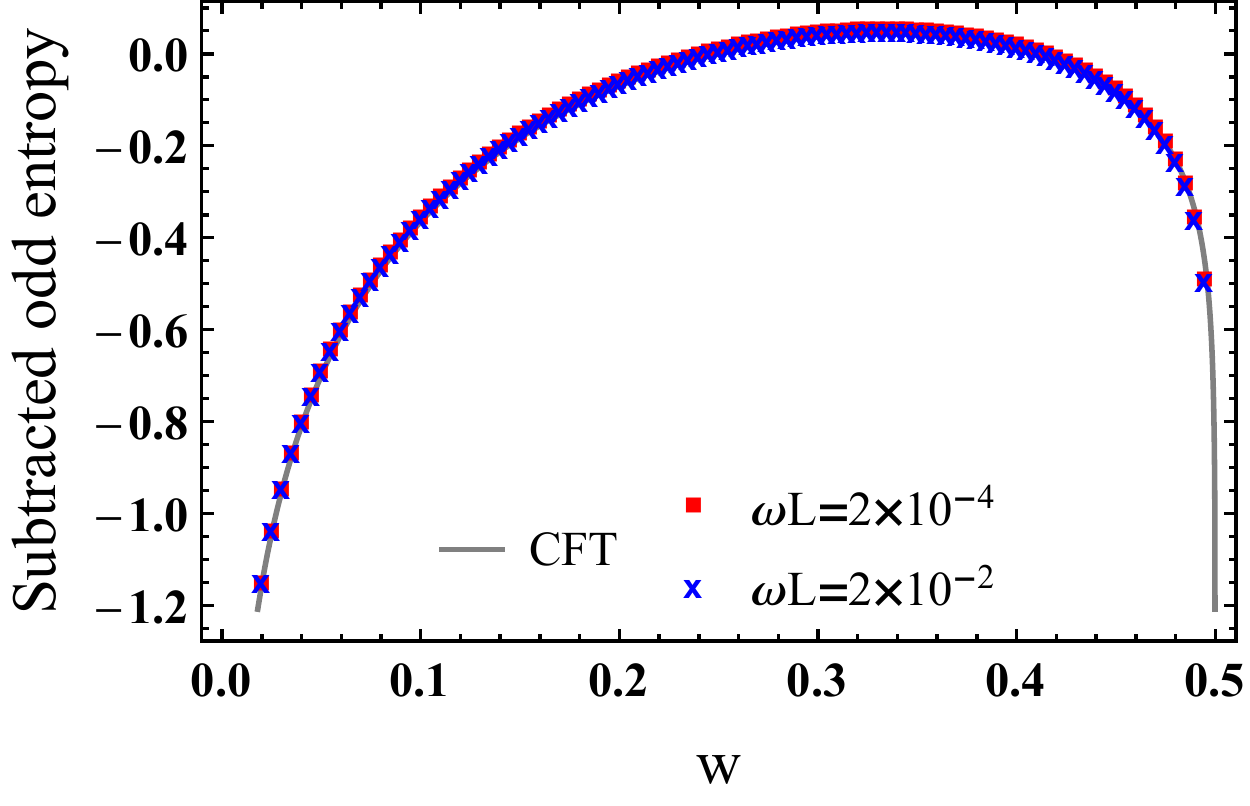}}
 \end{center}
 \caption{Odd entropy for two adjacent interval of equal length $\ell<L/2$ in a periodic chain with total size $L$. Here we defined $w=\ell/L$.}
  \label{fig:adjacent}
\end{figure}
First, we would like to compare our numerical result with the analytic one obtained from \eqref{eq:adj} so that we can confirm our result consistently reproduces the CFT result. 
In order to avoid lattice effect or cutoff parameter $\epsilon$ in \eqref{eq:adj}, we use the subtracted (or regularized) odd entropy defined as follows: 
\begin{align}
S^{\textrm{sub.}}_o(w)&=S_o(w)-S_o(1/4), \\
S_o(w)&=\dfrac{c}{6}\log\dfrac{\sin(\pi w)^2\sin(2\pi w)}{\e^3}+\textrm{const.}\,, 
\end{align}
where we defined $w\equiv\ell/L$. 
We also set $\ell_A=\ell_B\equiv\ell$ for simplicity and conformally mapped previous \eqref{eq:adj} to one on a cylinder with circumference $L$. In Figure \ref{fig:adjacent}, we plotted periodic chain results together with the subtracted odd entropy. It nicely gives the consistency between numerical and analytical results. Since the odd integer analytic continuation for analytic results potentially contain subtleties, this indeed gives a sanity check of replica trick calculations.

Note that the \eqref{eq:adj} is the same as the result in holographic CFTs. We can easily confirm this coincidence by using the replica trick for two-dimensional CFTs. This is because the result is determined by the three point function of the twist operators (in the two-dimensional case, these are just local operators). 

\subsection{Two disjoint interval}
\begin{figure}[t]
 \begin{center}
  \resizebox{160mm}{!}{
 \includegraphics{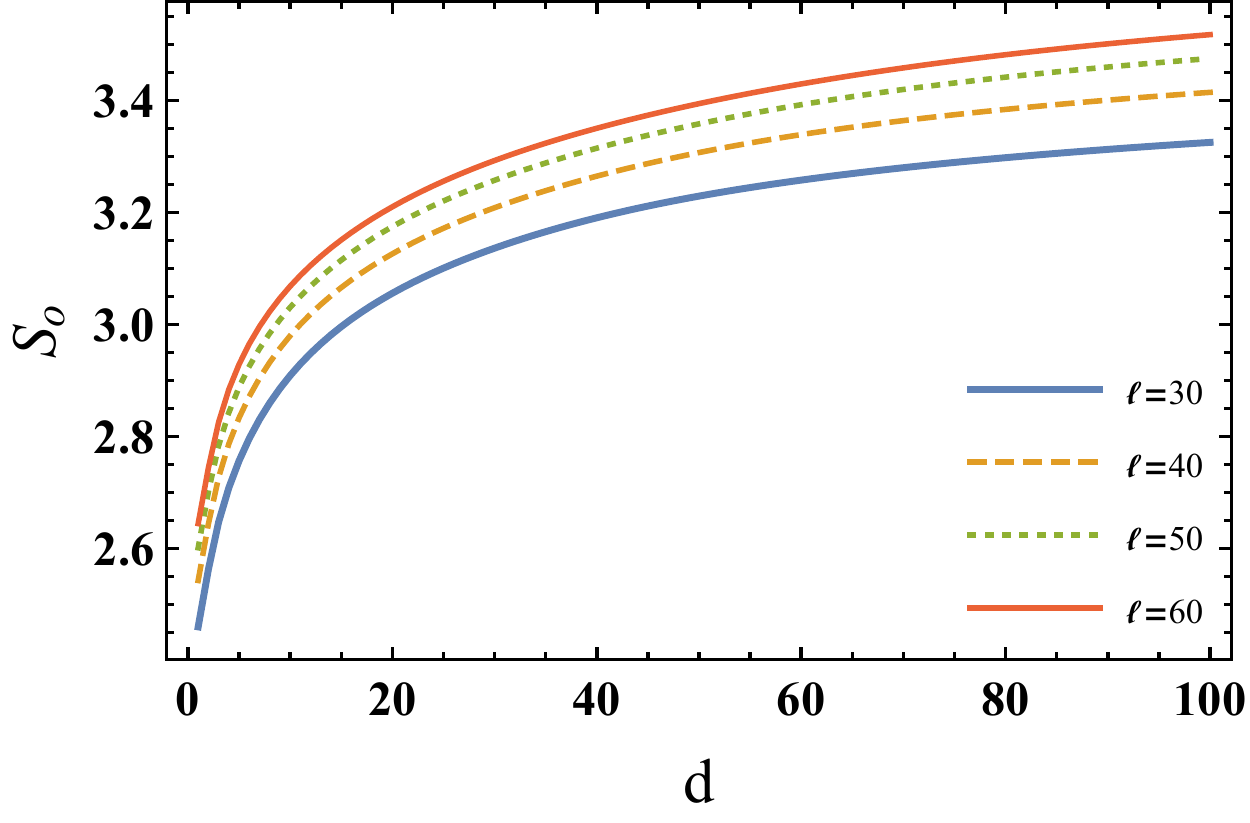}
 \includegraphics{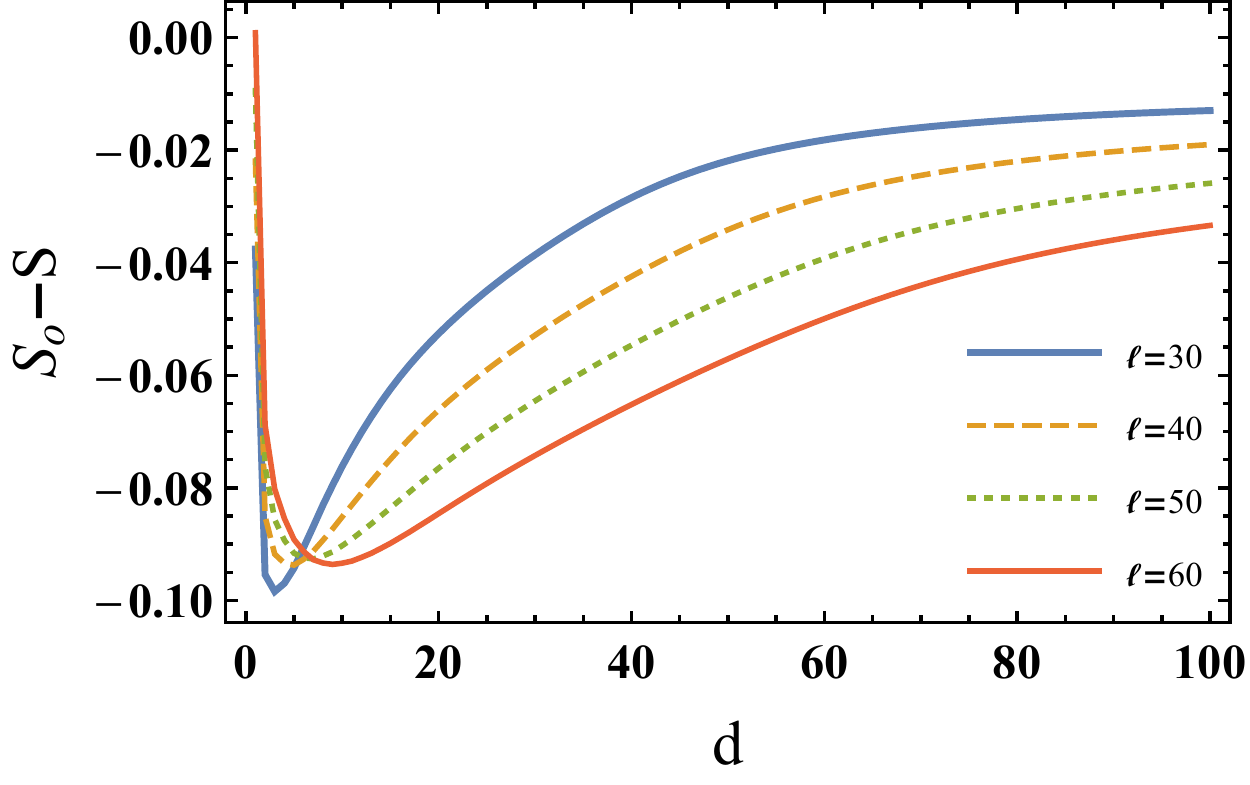}}
 \end{center}
 \caption{Left: a plot for $d$-dependence of the odd entropy. Right: a plot for $d$-dependence of the difference between odd entropy and von Neumann entropy. In both panels, we took $L=2000, \ell_A=\ell_B\equiv\ell$ and massless under the Dirichlet boundary condition. These figures already suggest that $S_o$ can be a monotonic function with respect to the subsystem size, whereas $S_o-S$ can not. }
 \label{fig:So_disjoint}
\end{figure}

Next, we move to disjoint interval cases ($d>0$ cases in Figure \ref{fig:setup}) where we have no full analytic expressions for now. We plotted it in Figure \ref{fig:So_disjoint}. As expected, $S_o$ is always positive. 

An interesting question is the behavior of odd entropy in small-$d$ and large-$d$ regions. 
We focus on the case $\ell_A=\ell_B=\ell$ for simplicity. For the small $d$, one can fit $S_o$ as a simple logarithmic function, 
\be
S_o=a_s\log\frac{d}{\ell}+b_s \;\;\;(d\ll\ell),
\ee
where $a_s$ and $b_s$ are positive constants. As we have already seen in the previous subsection \ref{subsec:vac_adj}, we could identify $a_s=1/2$ for strictly $d\rightarrow0$ limit, which is consistent with the CFT calculations. However, we have to report that for small but non-zero $d\ll\ell$, the coefficients $a_s$ and $b_s$ always depend on other parameters including the system size $L$. In such regimes, we cannot avoid lattice effects in general. 

For $d\gg\ell$, one can find 
\be
S_o=a_l \mathrm{e}^{-b_l\sqrt{\frac{d}{\ell}}}+c_l \;\;\;(d\gg\ell),
\ee
where $a_l,b_l,c_l$ are again positive constants. These constants also depend on other parameters including the system size $L$. This would be because the non-compactness of scalar fields. Fixing this behavior directly from other theories {\it e.g.} Ising model is an intriguing future direction. 

Next, we focus on the unusual shape of the curve in $S_o-S$ and discuss its origin. 
Although $S_o(A:B)$ behaves almost the same as von Neumann entropy $S(AB)$, the difference between them has a richer structure. In the middle distance, we have a small ``bump'' which makes $S_o-S$ a non-monotonic function of $d$. We can see that this bump becomes sharper if we increase classical correlations. We will be more concrete about this statement when we consider the same setup for finite-temperature cases. 

We should also notice that $S_o-S$ is \textit{not} a monotonic function for the subsystem size $\ell$. More precisely, we have a transition point at a certain value of $d$. Before and after this point, the direction of increase or decrease is reversed. We will discuss this non-monotonicity in more detail in section \ref{sec:ineq}. 

After the aforementioned ``bump'', we have an ``inflection point'' such that the $S_o-S$ can become closer to zero. The inflection point certainly appears when the negative eigenvalues of $\rho^{T_B}_{AB}$ effectively vanish. In other words, after the inflection point, classical correlations dominate. Thanks to the numerics, we can confirm this statement directly from the distribution of eigenvalues. One can also see a similar inflection point from a simple toy model of two-qubit states (see appendix \ref{app:twoq}). 

For sufficiently large $d$, $S_o-S$ reaches zero\footnote{Strictly speaking, our finite $L$ calculation suggests that we still hold small long-range entanglement. In other words, there is a tiny negative eigenvalue for $\rho^{T_B}_{AB}$. However, we report that this negative contribution becomes smaller and smaller as we increase the system size $L$.}. One may think our state becomes separable if we take large $d$ limit --- one important remark is that $S_o=S$ does not always imply $\rho_{AB}^{T_B}=\rho_{AB}$ in this limit. This caution is particularly important in our setup because there is a theorem that Gaussian states invariant under partial transposition are necessarily separable\cite{Lami2016}. For Lifshitz theories discussed in the next section, in specific cases, we observed $S_o-S=0$ at large $d$ together with $\rho^{T_B}_{AB}=\rho_{AB}$. 

\begin{figure}[h]
 \begin{center}
  \resizebox{160mm}{!}{
 \includegraphics{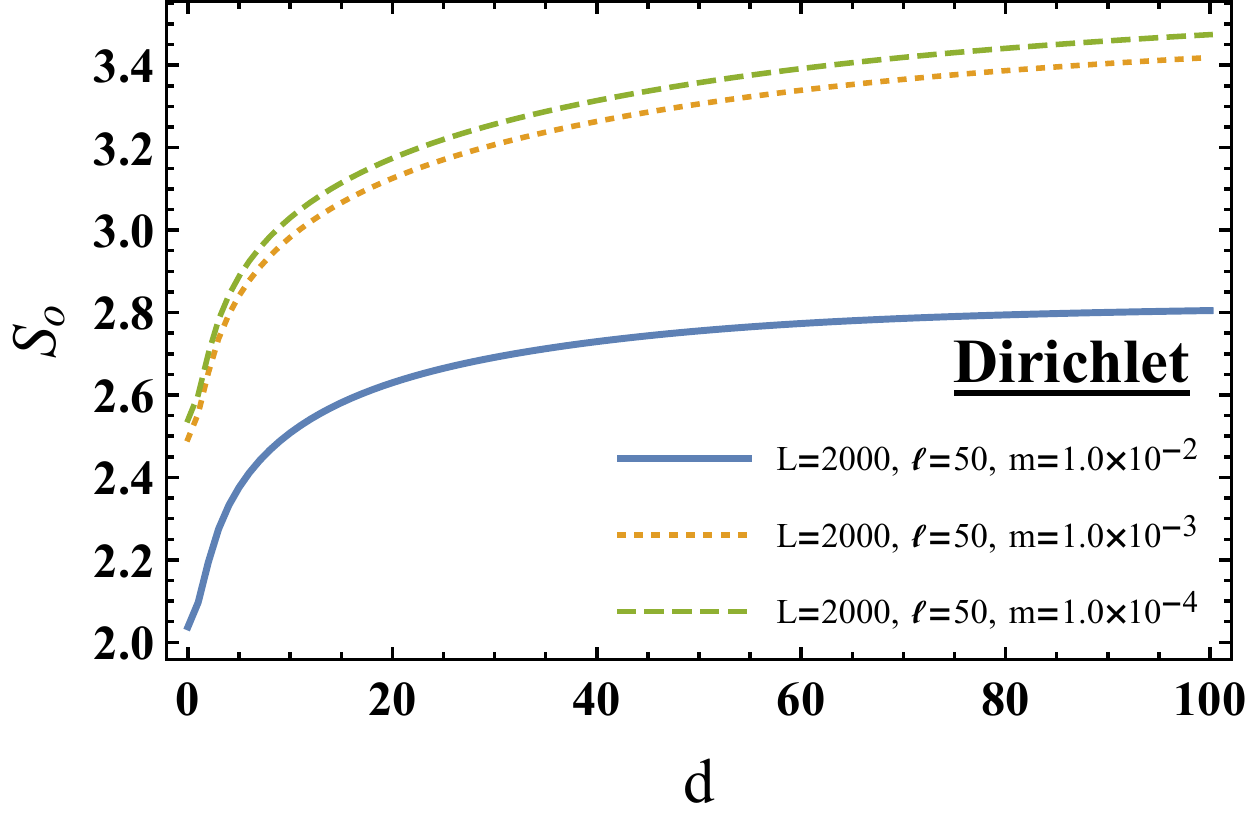}
 \includegraphics{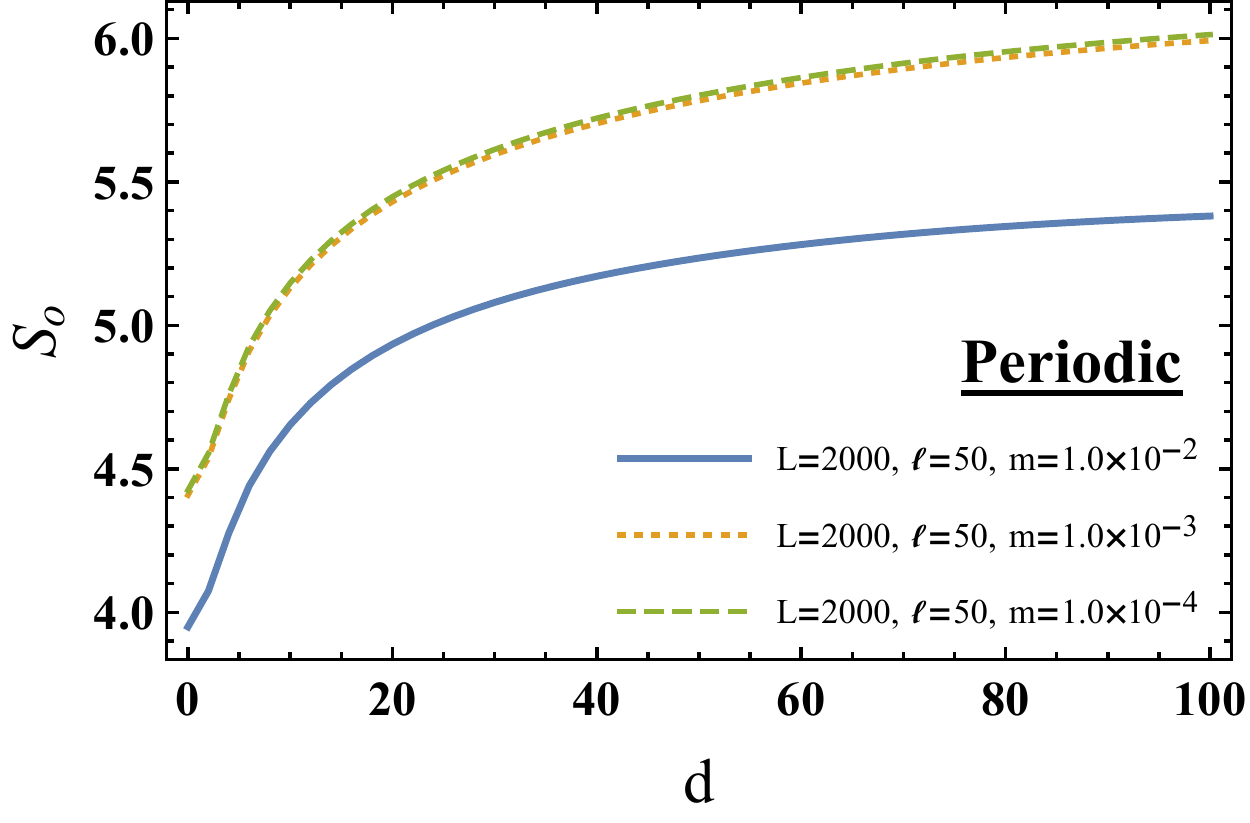}}
 \resizebox{160mm}{!}{
 \includegraphics{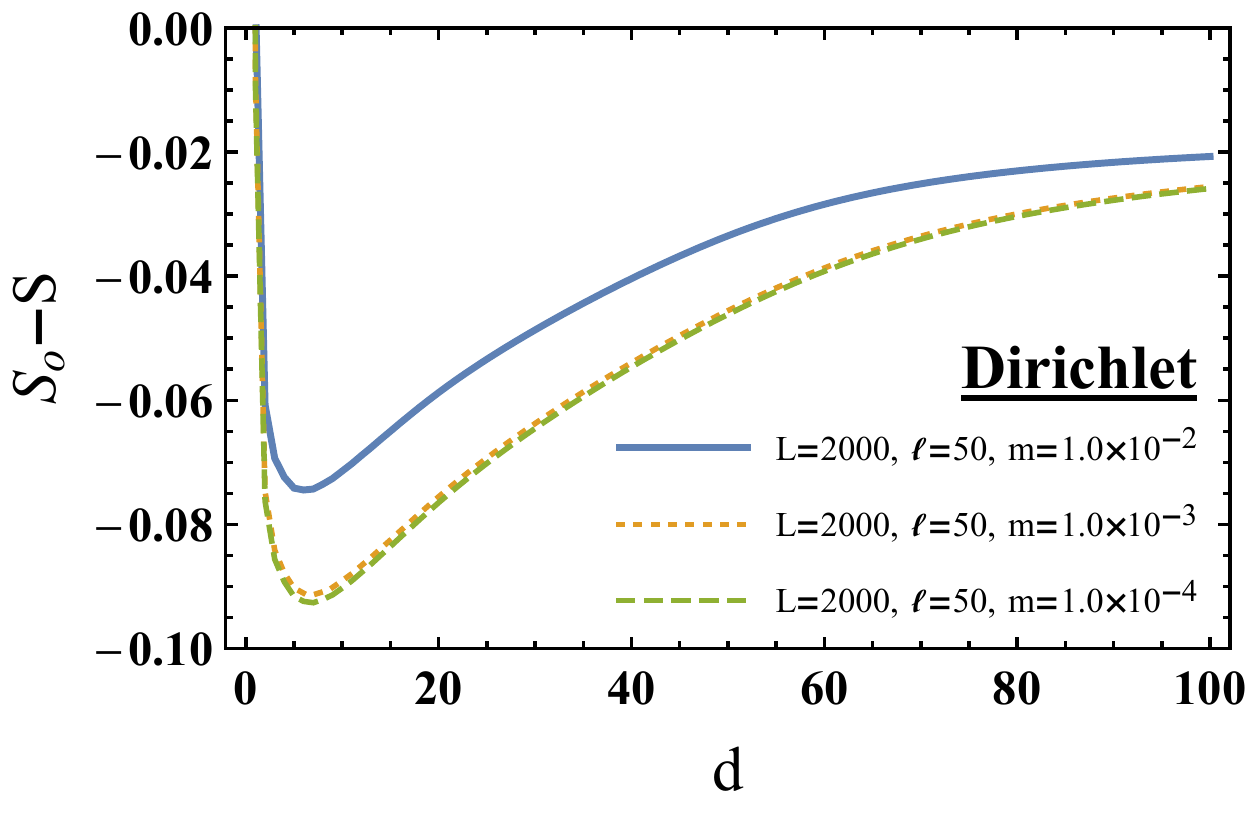}
 \includegraphics{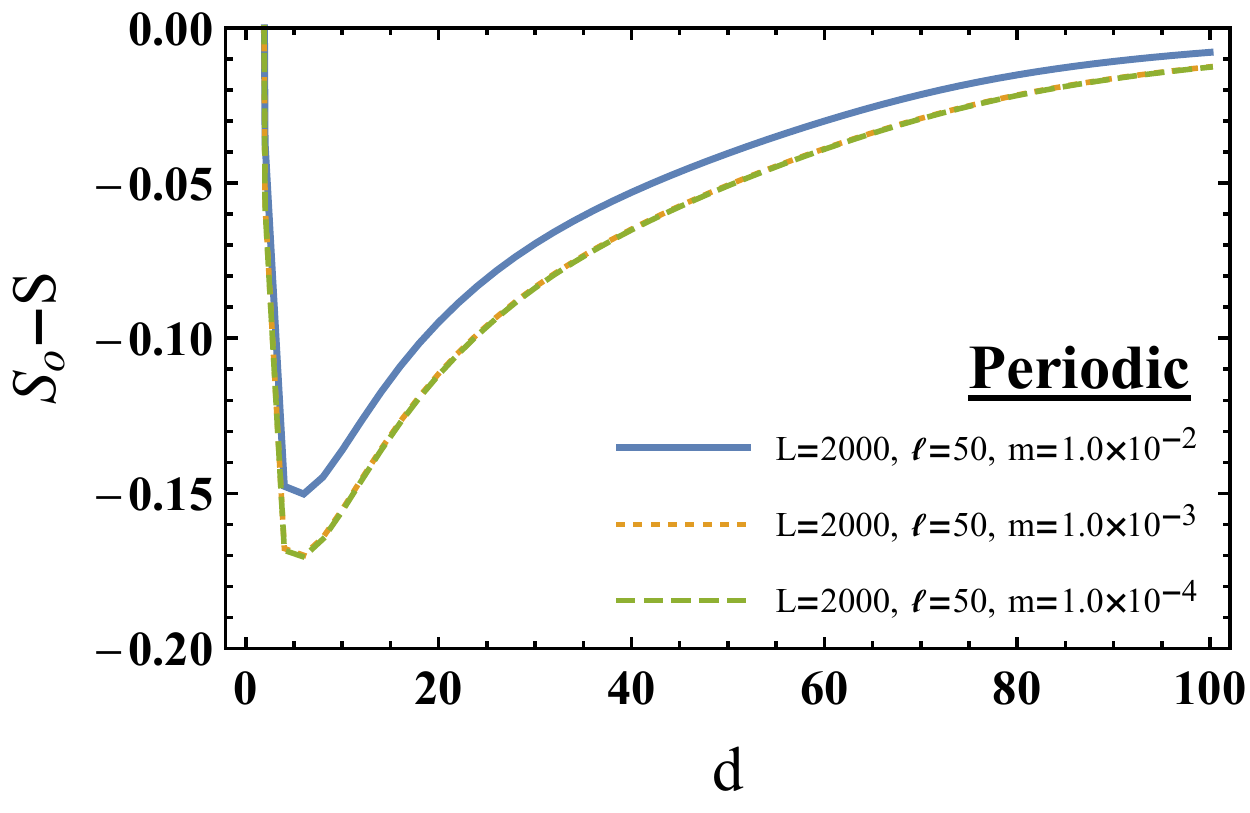}}
 \end{center}
 \caption{Mass and boundary condition dependence (left panels: Dirichlet, right panels: periodic) for odd entropy (upper panels) and deference between odd entropy and von Neumann entropy (lower panels). Here we took $L=2000, \ell_A=\ell_B=50$. }
 \label{fig:dp_mass}
\end{figure}
\subsection{Comments on mass-dependence and boundary conditions}

So far we mostly discussed massless scalar fields with Dirichlet boundary conditions. 
Here we briefly comment on the dependence on the mass parameter $m$\footnote{From now, we will use $m$ instead of $\omega$ as a mass parameter.} and to boundary conditions (Dirichlet or periodic). Figure \ref{fig:dp_mass} shows the mass and boundary condition dependence of $S_o$ and $S_o-S$. As we can see, there is no qualitative difference between Dirichlet and periodic boundary conditions except for the long-range behavior. 

For the Dirichlet boundary condition, we still have small deference between $S_o$ and $S$ even for large $d$ limit. We expect this is just a boundary effect. As we have already mentioned, this deviation is controllably suppressed as we increase the total subsystem size $L$. Figures for massive cases may suggest that we should consider absolute value if one wants to regard $S_o-S$ as a measure of certain correlations. 

\section{Classical versus quantum correlations in odd entropy}
\subsection{Increase classical correlations: thermal states}\label{subsec:thm}
Next we consider thermal states which are genuinely mixed states. A motivation to study these states is to confirm the sensitivity of odd entropy against classical correlations. We are particularly interested in the behavior of odd entropy ($S_o$) and the subtracted version of it ($S_o-S$) under the increase of temperature (classical correlations) in the original density matrix. 

\begin{figure}[h]
 \begin{center}
  \resizebox{80mm}{!}{
  \includegraphics{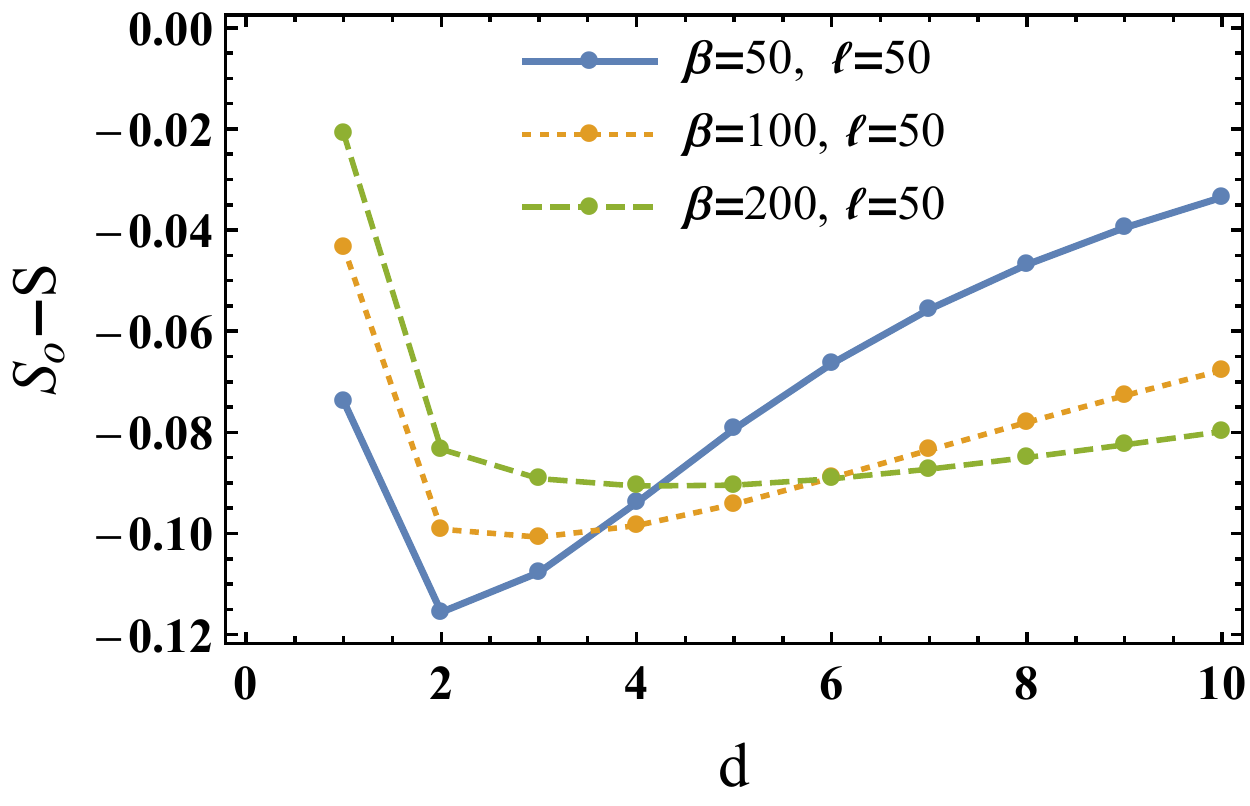}
 }\resizebox{88mm}{!}{
  \includegraphics{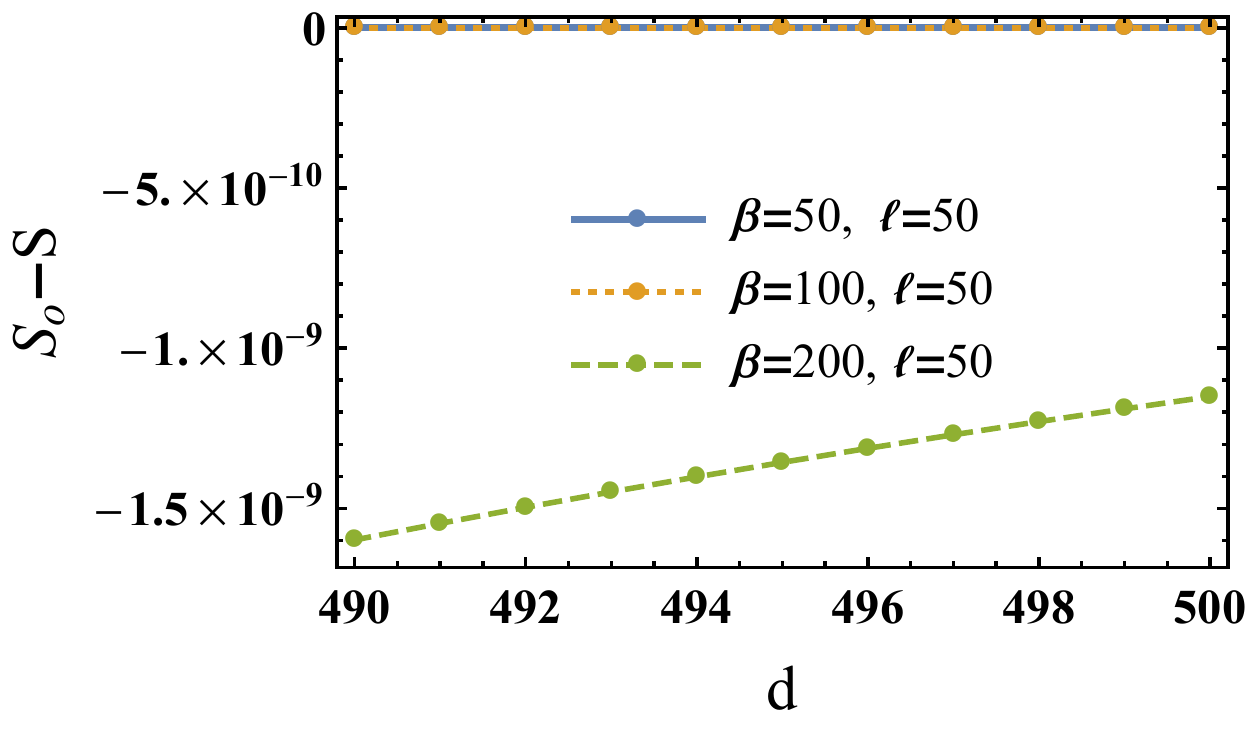}
 }
 \end{center}
 \caption{Behaviors of $S_o-S$ at finite temperature $\beta^{-1}$ with the fixed subsystem size $\ell_A=\ell_B\equiv\ell$. Here we took $L=2000$. Left: small-$d$ regime. Right: large-$d$ regime. The results for $\beta=50$ and $100$ are degenerated. }
 \label{fig:thm2_1}
\end{figure}

The whole shape of the figure, in this case, is qualitatively similar to the one in the vacuum. In particular, we still have the bump in the middle range of $d$ and can learn how it is sensitive to the temperature (see right panel of Figure \ref{fig:thm2_1}). Clearly, the minimum value becomes smaller as we increase the temperature. It means that the strange bump in $S_o-S$ is sensitive to the classical correlations. 

On the other hand, in large $d$ regime, we observed that the $S_o-S$ reaches to strictly zero as opposed to the vacuum case with finite $L$ (see right panel of Figure \ref{fig:thm2_1}). Therefore, in this regime, we expect that $S_o-S$ does not pick up even classical correlations. 

To summarize, if we have both quantum and classical correlations, the behavior of $S_o-S$ is determined from both of them, however, once classical correlations become dominant, $S_o-S$ becomes insensitive even to the classical correlations. 

\subsection{Increase quantum correlations: Lifshitz theories}\label{subsec:lif}
So far we have described Gaussian states in the discretized version of a relativistic scale-invariant model. We can also extend our study to discretized counterparts of non-relativistic scale-invariant theories, known as Lifshitz theories. These theories, as well as their harmonic lattice realization, are invariant under anisotropic scaling between the spacial and temporal directions. This anisotropy is parametrized by a dynamical critical exponent denoted by $z$. 

Interestingly, as we increase the value of $z$, the system acquires longer-range total correlations. In particular, if we focus on the short distance regime, we can expect such total correlations are dominated by quantum ones\cite{MohammadiMozaffar:2017chk}. Therefore, in contrast to the thermal states, the $z$-dependence of Lifshitz theories tells us how odd entropy is sensitive to such quantum correlations. 

\begin{figure}[h]
 \begin{center}
 \resizebox{160mm}{!}{
 \includegraphics{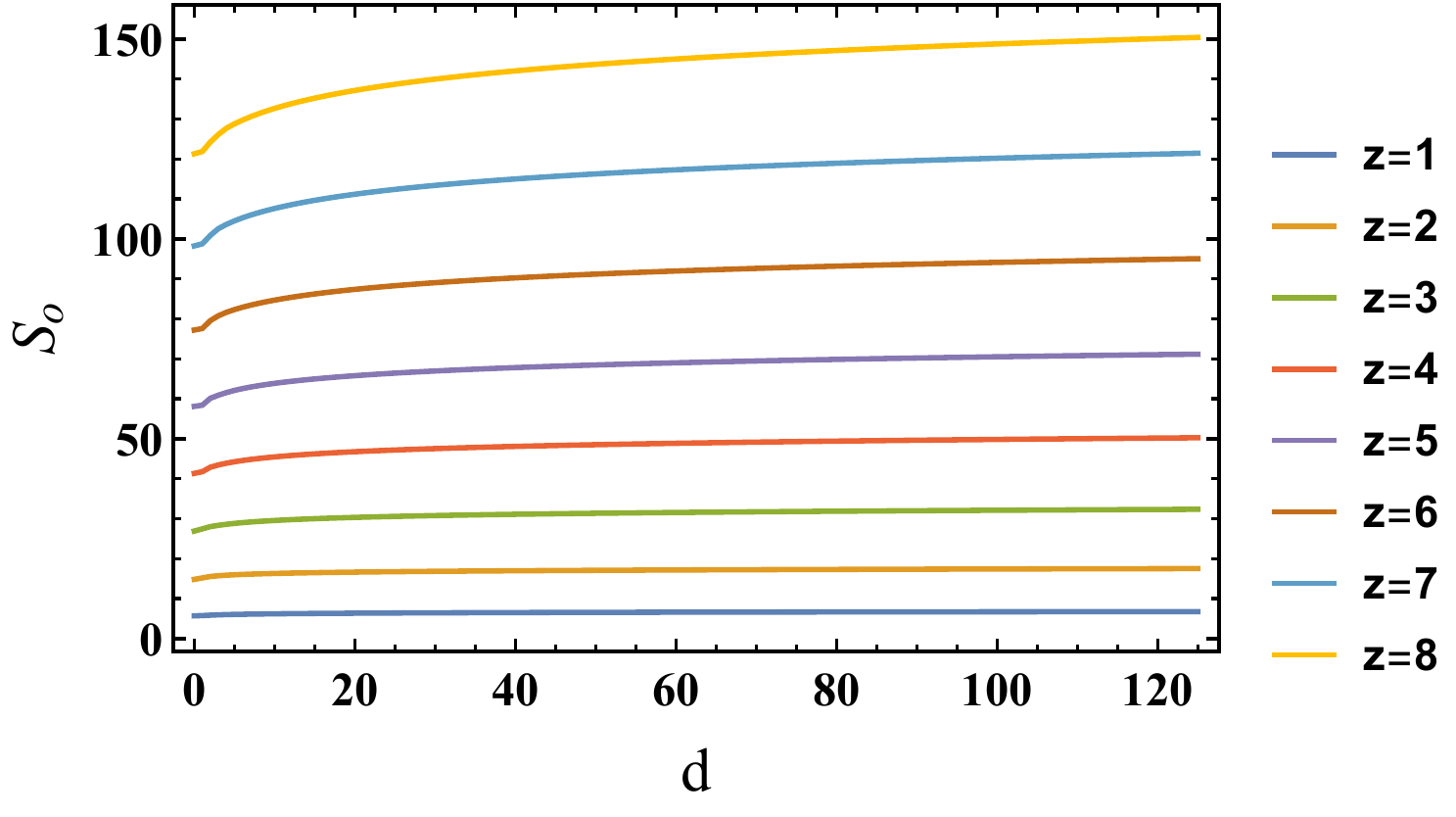} \includegraphics{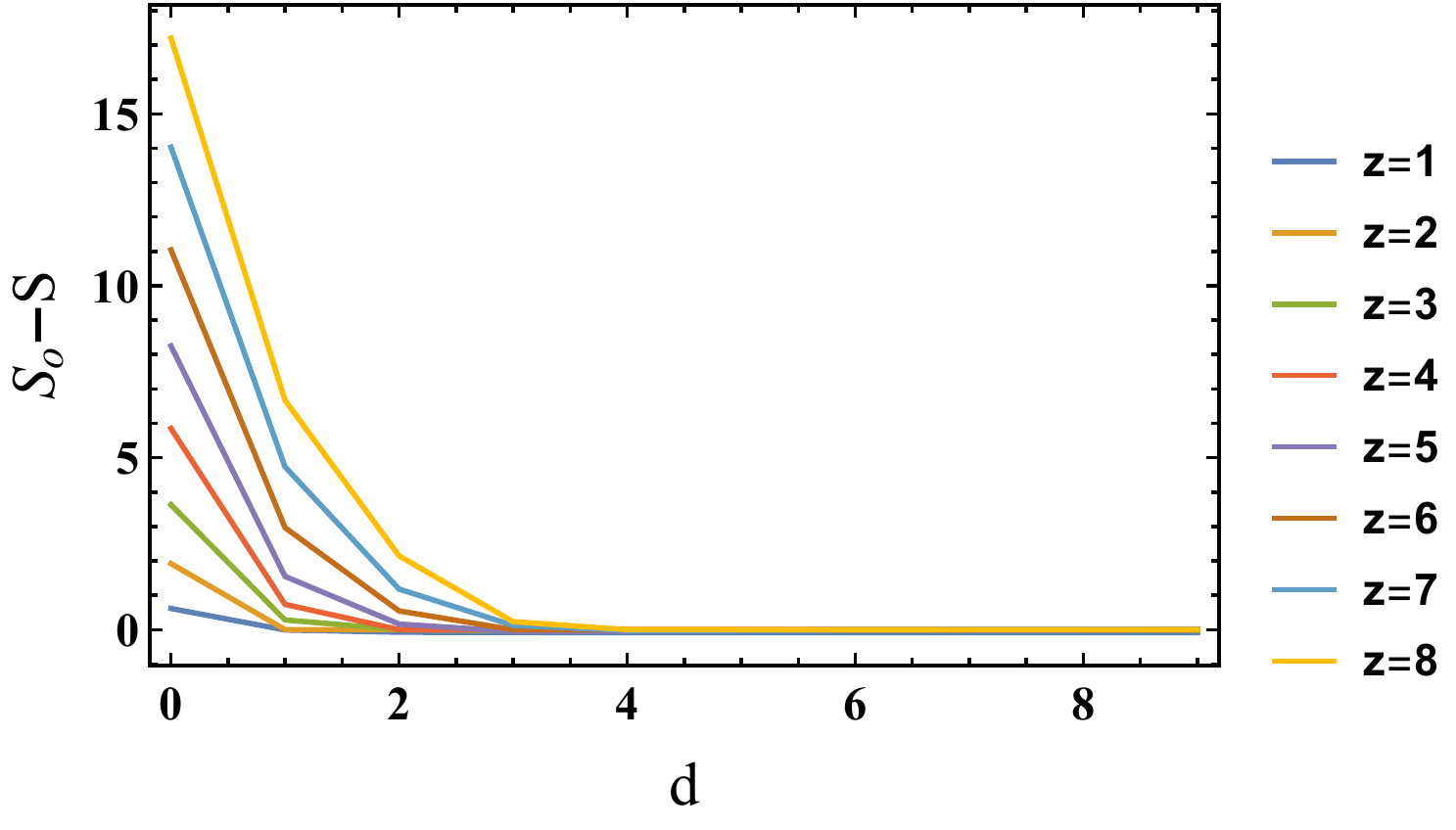}}
   \resizebox{160mm}{!}{
  \includegraphics{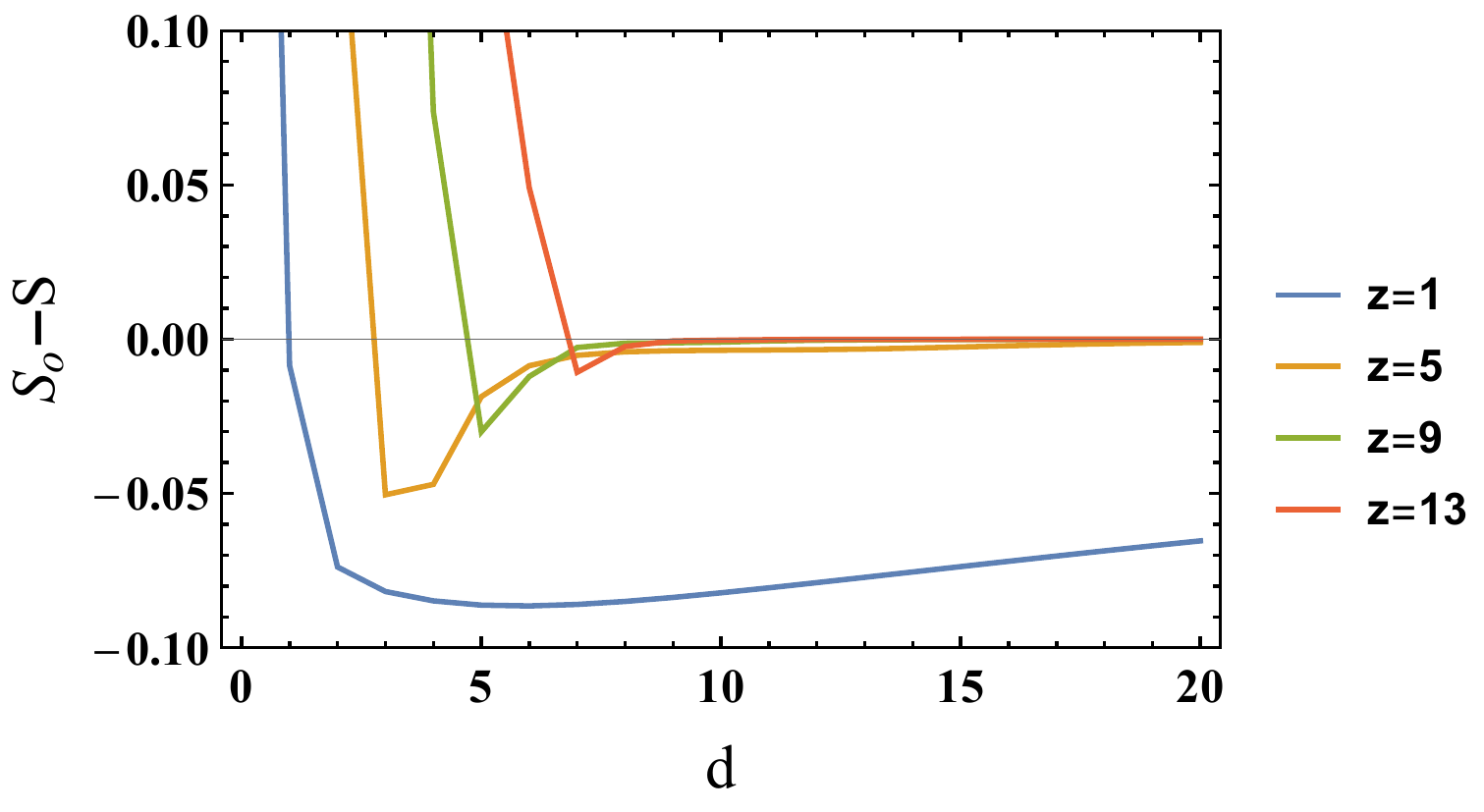} \includegraphics{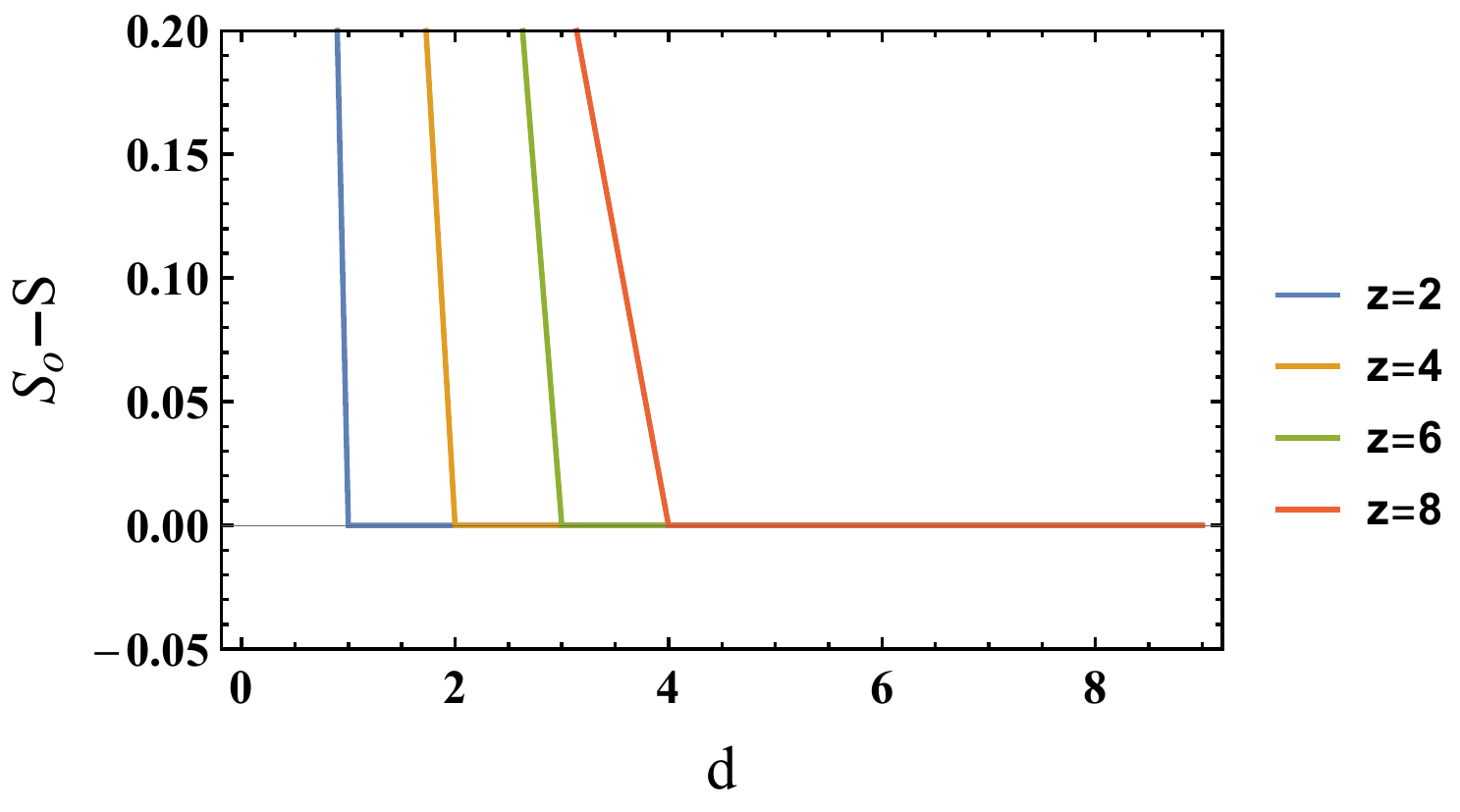}}
 \end{center}
 \caption{Left: $d$-dependence of the odd entropy for various $z$. Right: one of the difference between odd entropy and von Neumann entropy. In both panels, we took $L=2000, \ell_A=\ell_B\equiv\ell$ and $m=1.0\times 10^{-6}$ under the periodic boundary condition. (Again, there are no difference between this and massless Dirichlet boundary condition.) Notice that the minimum of $S_o-S$ increases for $z>1$. }
 \label{fig:oddz}
\end{figure}

We plotted $d$-dependence of the $S_o$ and $S_o-S$ in Figure \ref{fig:oddz}. First, let us look at the small-$d$ regime (upper panels in Figure \ref{fig:oddz}). 
Clearly, both of them increase as we increase the value of $z$. This is quite reasonable from the harmonic chain Hamiltonian which acquires longer range correlations as we increase the value of $z$. 

Next, let us zoom up around the minimum value of $S_o-S$ (lower panels in Figure \ref{fig:oddz}). In there, we can classify the shape of $S_o-S$ with the parity of $z$. For odd $z$, we can find similar behavior as $z=1$ case as we have seen in section \ref{sec:vac}. One remarkable point is that the minimum value of $S_o-S$ increases as we increase the value of $z$. 

On the other hand, for even $z$, it is always positive semidefinite and suddenly decays to zero\footnote{See also \cite{Chen:2017txi} where similar behavior has been reported for specific models with $z=2$.}. Interestingly, this is completely the same behavior as the logarithmic negativity, 
\be
\mathcal{E}_{LN}(A:B)=\log|\rho_{AB}^{T_B}|_1=\log\left(1+2\sum_{\lambda_i<0}|\lambda_i|\right),
\ee
which quantifies only quantum correlations\footnote{See also recent interesting work for Lifshitz theories\cite{Angel-Ramelli:2020wfo}. In there, we can see the coincidence of $S_o-S$ and the logarithmic negativity in some specific cases including higher dimension analytically.}. Here $|X|_1\equiv\mathrm{Tr}\sqrt{XX^\dagger}$. 
We have to note that in this regime, the mutual information,
\be
I(A:B)=S(A)+S(B)-S(AB),
\ee
still takes non-zero value and behaves as a monotonically decreasing function with respect to $d$. It means that for Lifshitz theories with even $z$, the $S_o-S$ counts only quantum correlations. For odd values of $z$, the larger the value of $z$, the closer gets $S_o-S$ to the logarithmic negativity. We leave numerical plots of these measures in appendix \ref{app:MI}.

\section{Some inequalities for odd entropy}\label{sec:ineq}
\subsection{Monotonicity}
\begin{figure}[h]
 \begin{center}
  \resizebox{160mm}{!}{
 \includegraphics{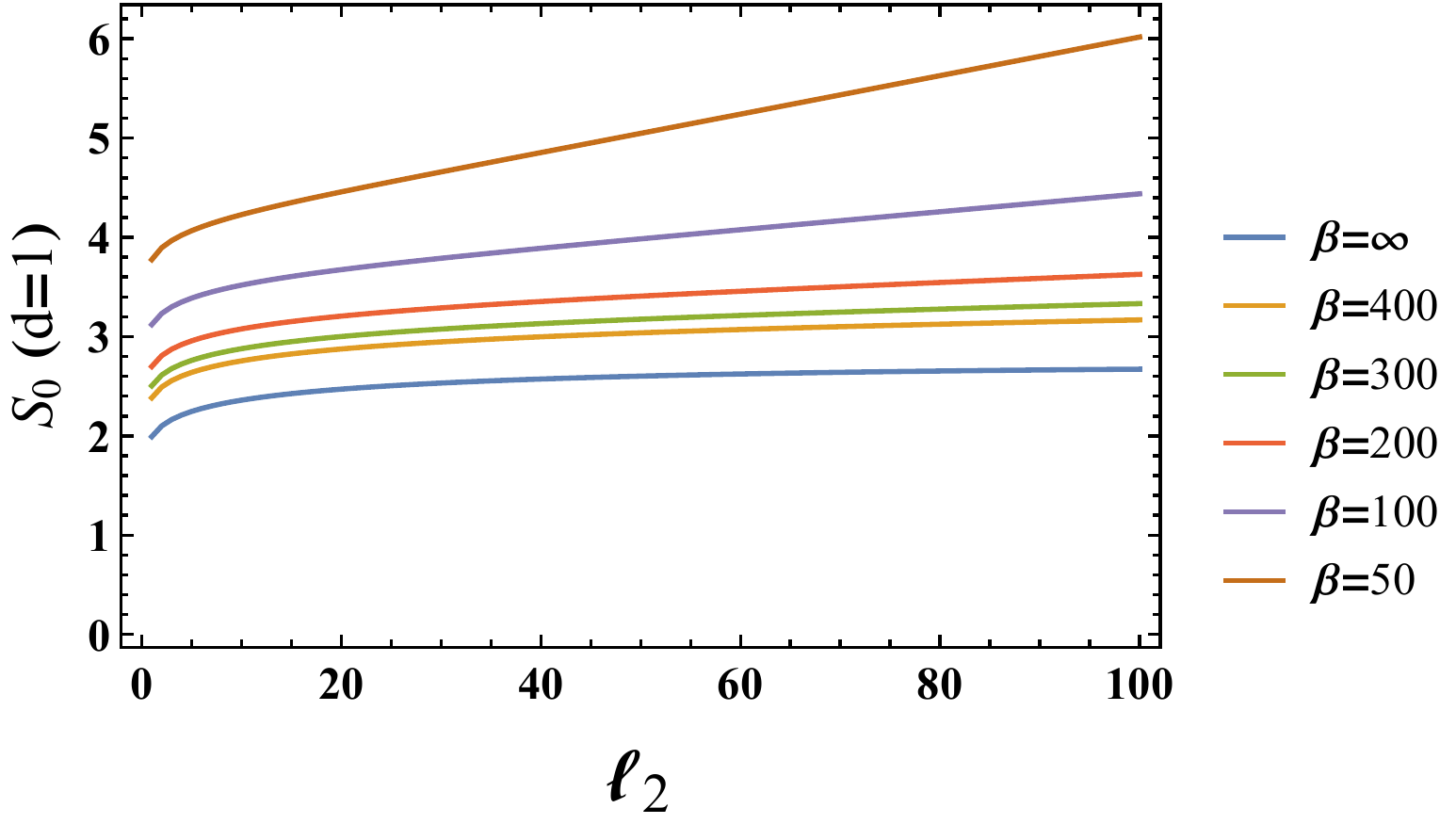}\hspace{1cm}\includegraphics{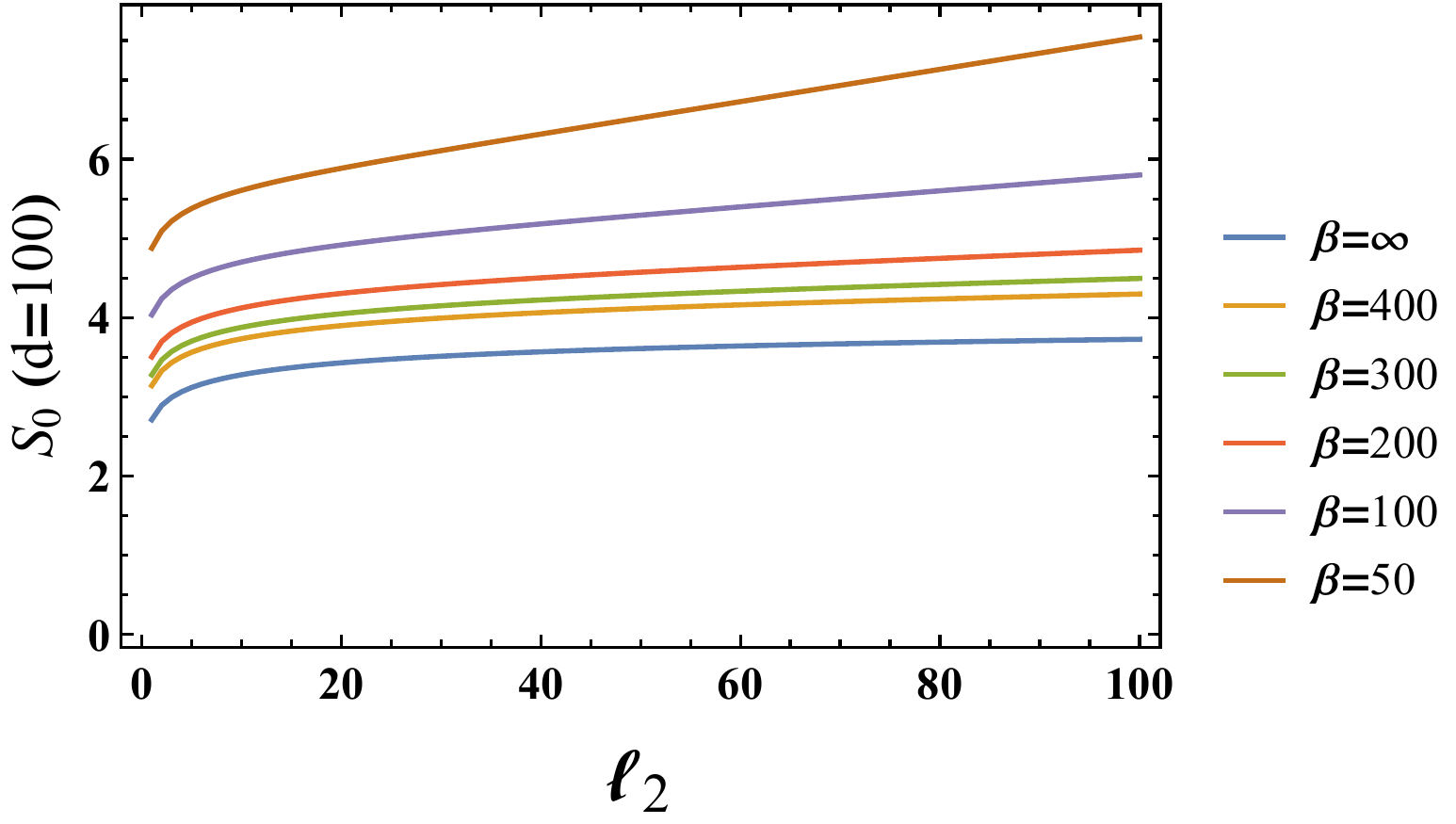}
 }\\
   \resizebox{160mm}{!}{
 \includegraphics{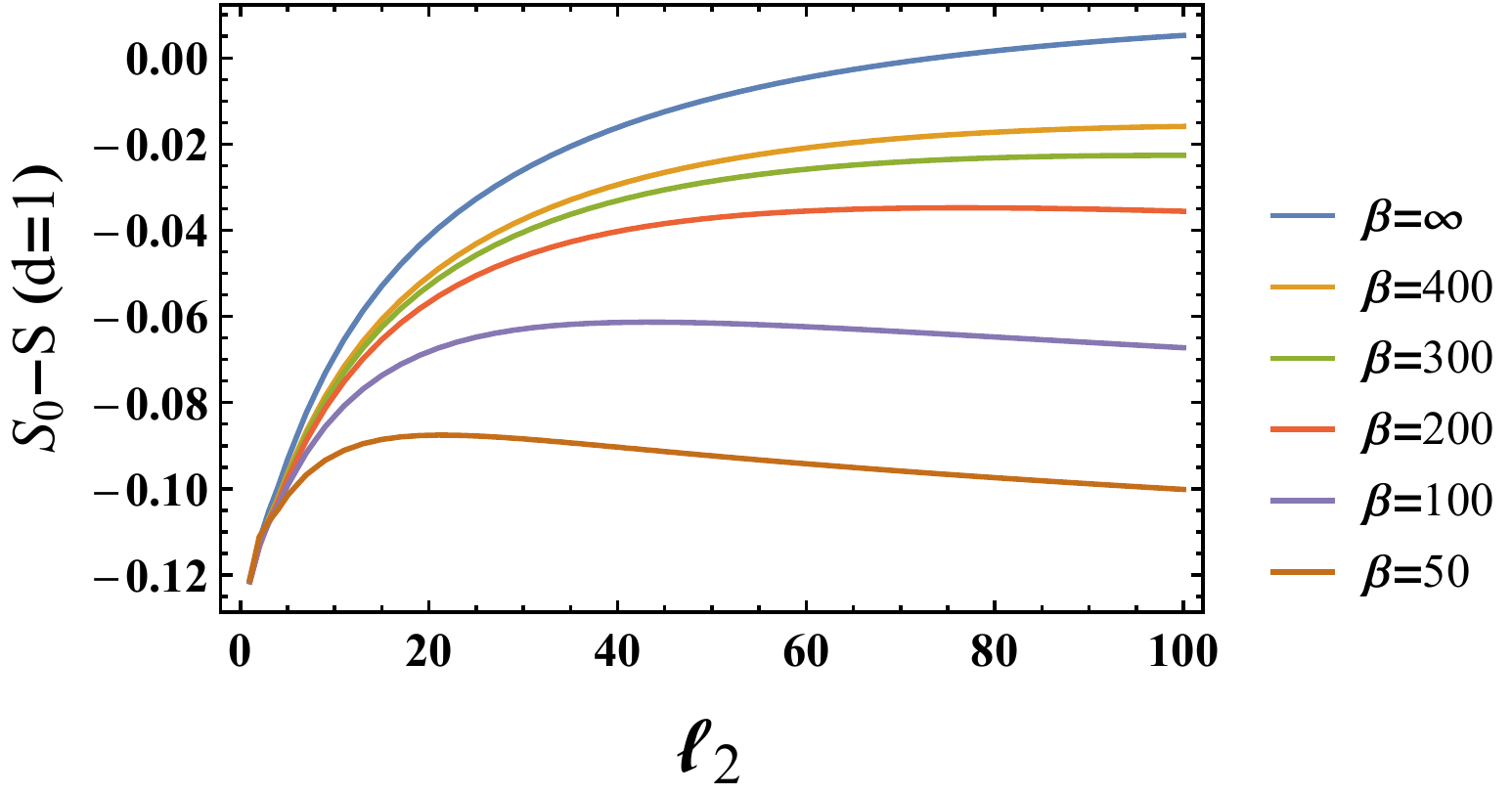}\hspace{1cm}\includegraphics{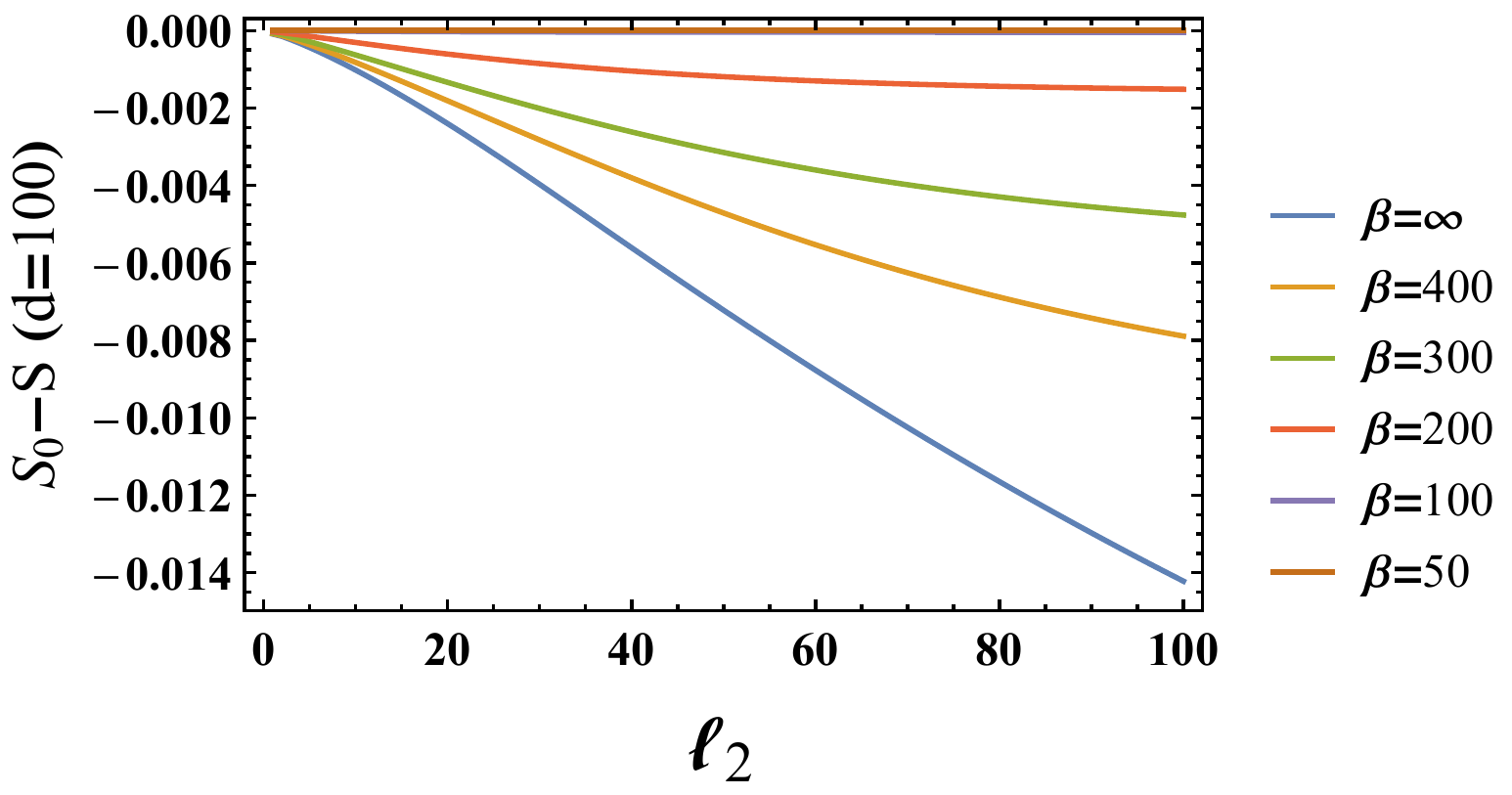}
 }
 \end{center}
 \caption{Monotonicity plots for $S_o$ and $S_o-S$. Here we took $L=2000, \ell_A=50$ and $d=1,100$ (short and long distance regime) with massless Dirichlet chain. We plotted $\ell_2\equiv\ell_B$ dependence. Figure suggests that finite temperature effects (increase of classical correlations) do break the monotonicity even for fixed $d$. Interestingly, the short and long distance behaviors of $S_o-S$ reflect monotonicity of (almost) opposite direction. }
 \label{fig:mono}
\end{figure}

For any measures of correlations $\mathcal{E}$, we should have the so-called monotonicity relation,
\be
\mathcal{E}(A:B_1B_2)\geq \mathcal{E}(A:B_1). \label{eq:monotonicity}
\ee
It is a natural requirement since enlarging one of two subsystems should increase the total amounts of correlations sharing between the two subsystems. We will check the relation \eqref{eq:monotonicity} in the case of $\mathcal{E}=S_o$ and $\mathcal{E}=S_o-S$. The Figure \ref{fig:mono} shows that $S_o$ nicely satisfies monotonicity \eqref{eq:monotonicity}, whereas $S_o-S$ does not. It means that $S_o-S$ cannot be interpreted as a correlation measure in general\footnote{We are able to define some measures from $S_o-S$ which satisfy the monotonicity. A trivial modification might be $\mathrm{max}\{S_o-S,0\}$ although it behaves almost everywhere trivially in our setup. }. From the holographic perspective, this is rather surprising because the entanglement wedge cross section clearly satisfies the monotonicity. We can also confirm the monotonicity for $\mathcal{E}=S_o$ and breaking for $\mathcal{E}=S_o-S$ with $z>1$. 

\subsection{Violation of monogamy relation and strong superadditivity}
\begin{figure}[h]
 \begin{center}
   \resizebox{160mm}{!}{
 \includegraphics{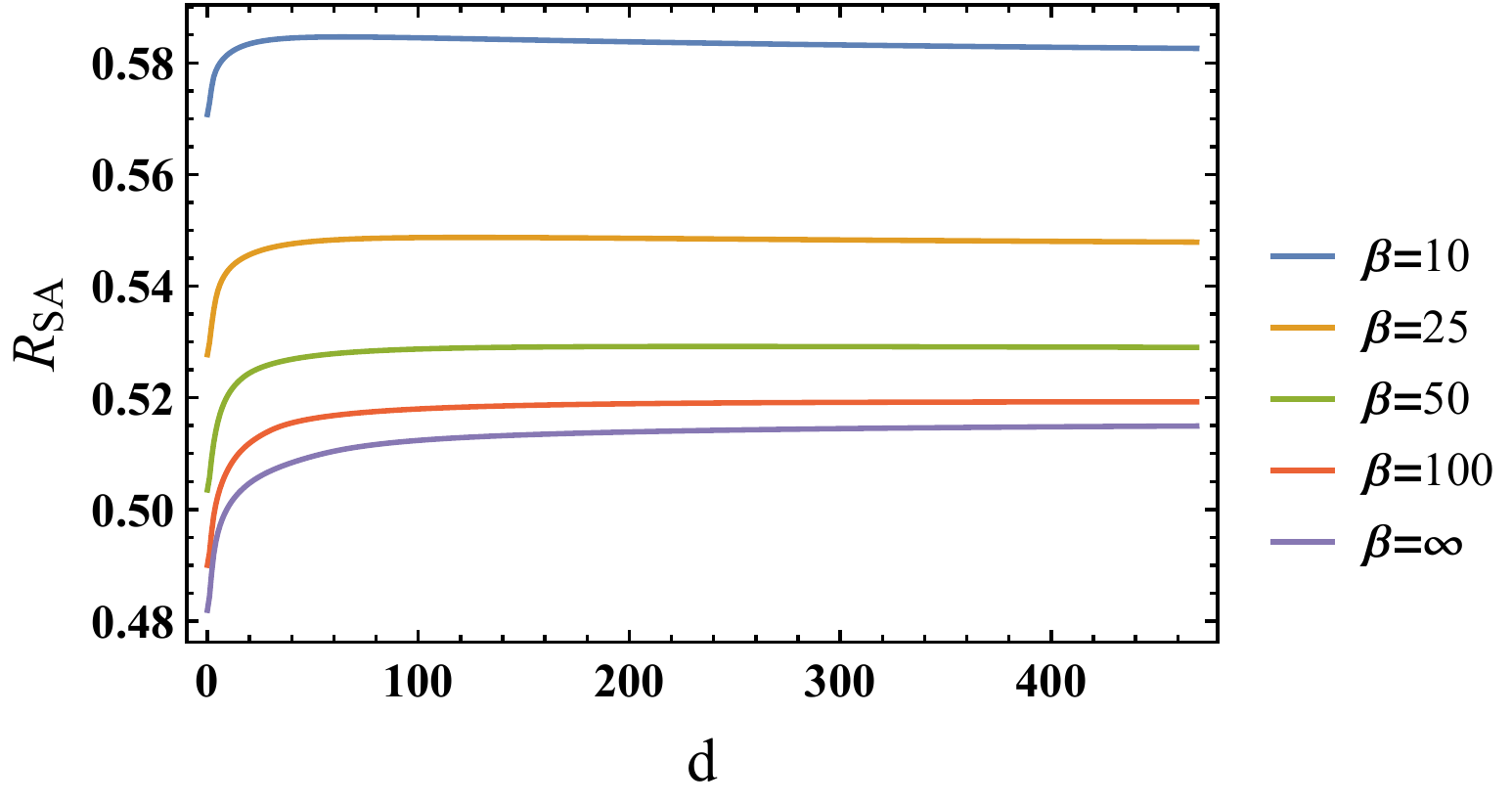}
\includegraphics{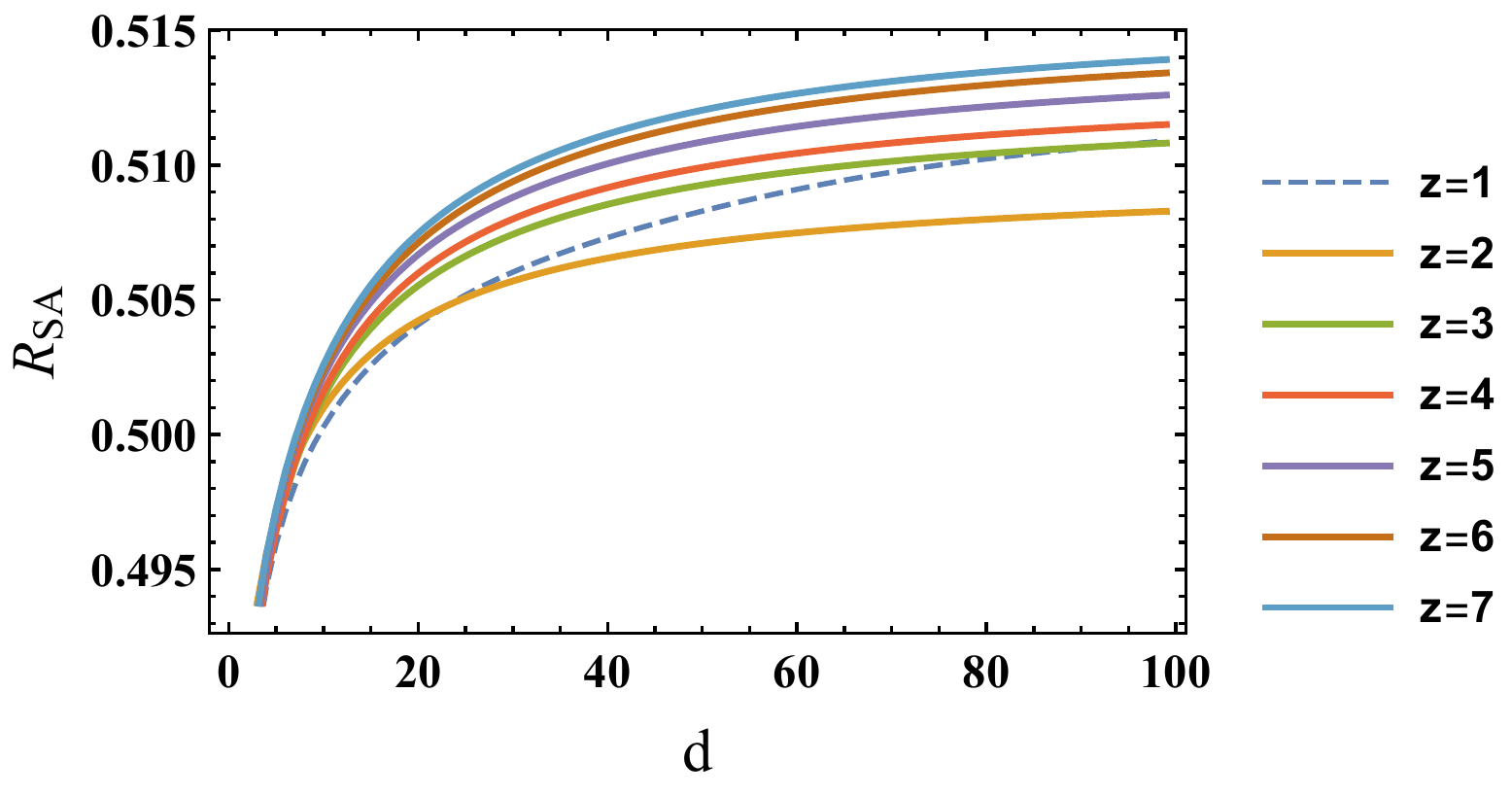}}
 \end{center}
 \caption{Plot for a ratio which suggests the polygamy relation of the odd entropy for the vacuum state. Here we took $L=2000, \ell_A=50, \ell_{B_1}=\ell_{B_2}=25$ with the periodic boundary condition. In these figures, we took $m=1.0\times 10^{-4}$. Left: (Inverse) temperature dependence of the same ratio as the left panel. Right: $z$-dependence of the same ratio as the left panels.  }
 \label{fig:rsa} 
\end{figure}
We expect measures of quantum correlations $\mathcal{E}$ to satisfy the so-called monogamy relation\cite{KoashiWinter, CO}, 
\be
\mathcal{E}(A:B_1B_2)\geq \mathcal{E}(A:B_1)+\mathcal{E}(A:B_2). \label{eq:monogamy}
\ee
For example, squashed entanglement, an ideal axiomatic measure of bi-partite quantum entanglement for mixed states, always satisfies this relation. On the other hand, mutual information, for example does not always satisfy it. In particular, we know that the \eqref{eq:monogamy} for mutual information is violated for our free scalar fields\cite{Casini:2008wt}. 
The opposite inequality is called as the polygamy relation. We expect quantum entanglement to be monogamous, whereas inclusion of (a certain amounts of) classical correlations lead the opposite (polygamy) relation. 
We also have a relation for a 4-partite state $\rho_{A_1A_2B_1B_2}$, strong superadditivity,
\be
\mathcal{E}(A_1A_2:B_1B_2)\geq \mathcal{E}(A_1:B_1)+\mathcal{E}(A_2:B_2). \label{eq:ssa}
\ee

Here we would like to ask whether $S_o$ in our scalar field theories can satisfy the monogamy relation \eqref{eq:monogamy} and the strong superadditivity \eqref{eq:ssa}. To investigate these properties, let us define
\be
R_{SA}=\dfrac{S(A:B_1B_2)}{S(A:B_1)+S(A:B_1B_2)},
\ee

The Figures \ref{fig:rsa} shows $S_o$ does always break the monogamy relation. It means that $S_o$ measures not only quantum correlations but also classical ones very much. This is an expected feature from our previous results. Since $S_o$ appears to be an entropy, we may call this relation as ``subadditivity'' of odd entropy rather than the polygamy relation. 

\begin{figure}[h]
 \begin{center}
  \resizebox{140mm}{!}{\includegraphics{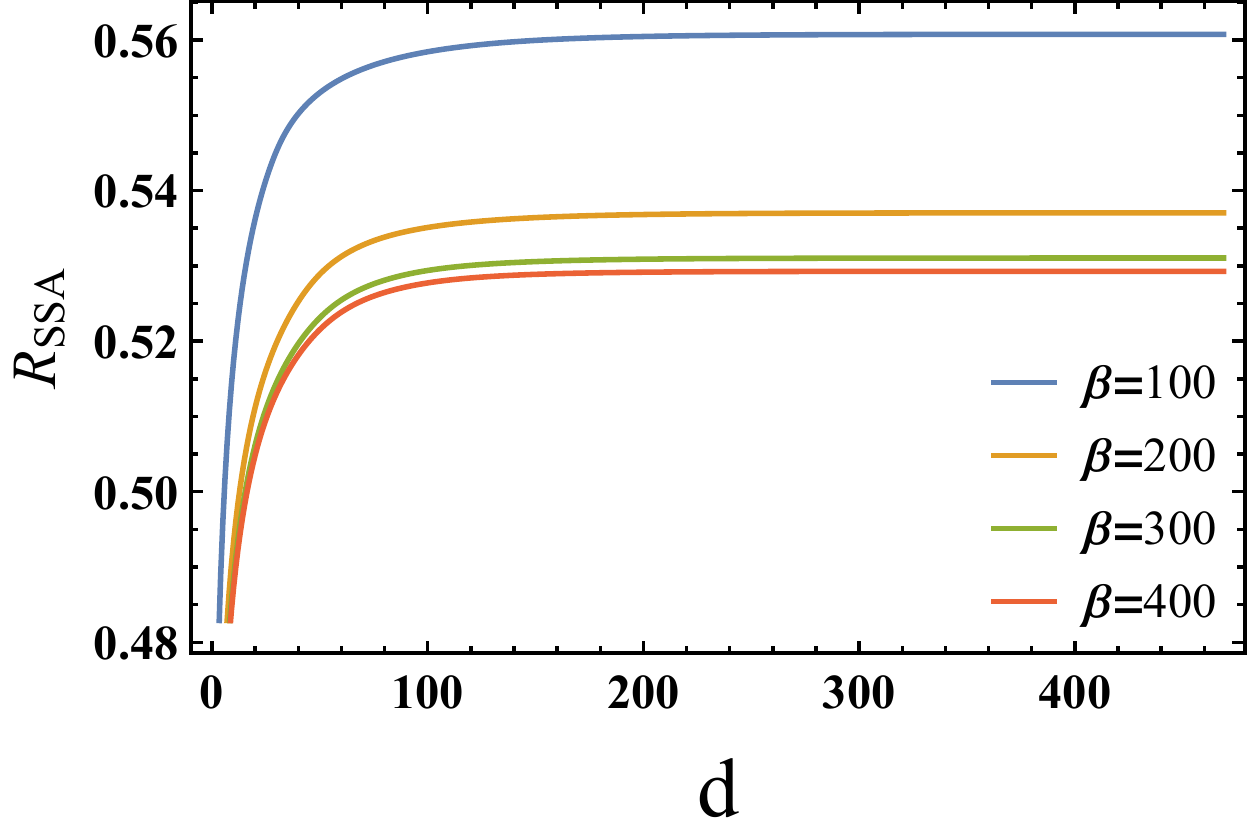}
 \includegraphics{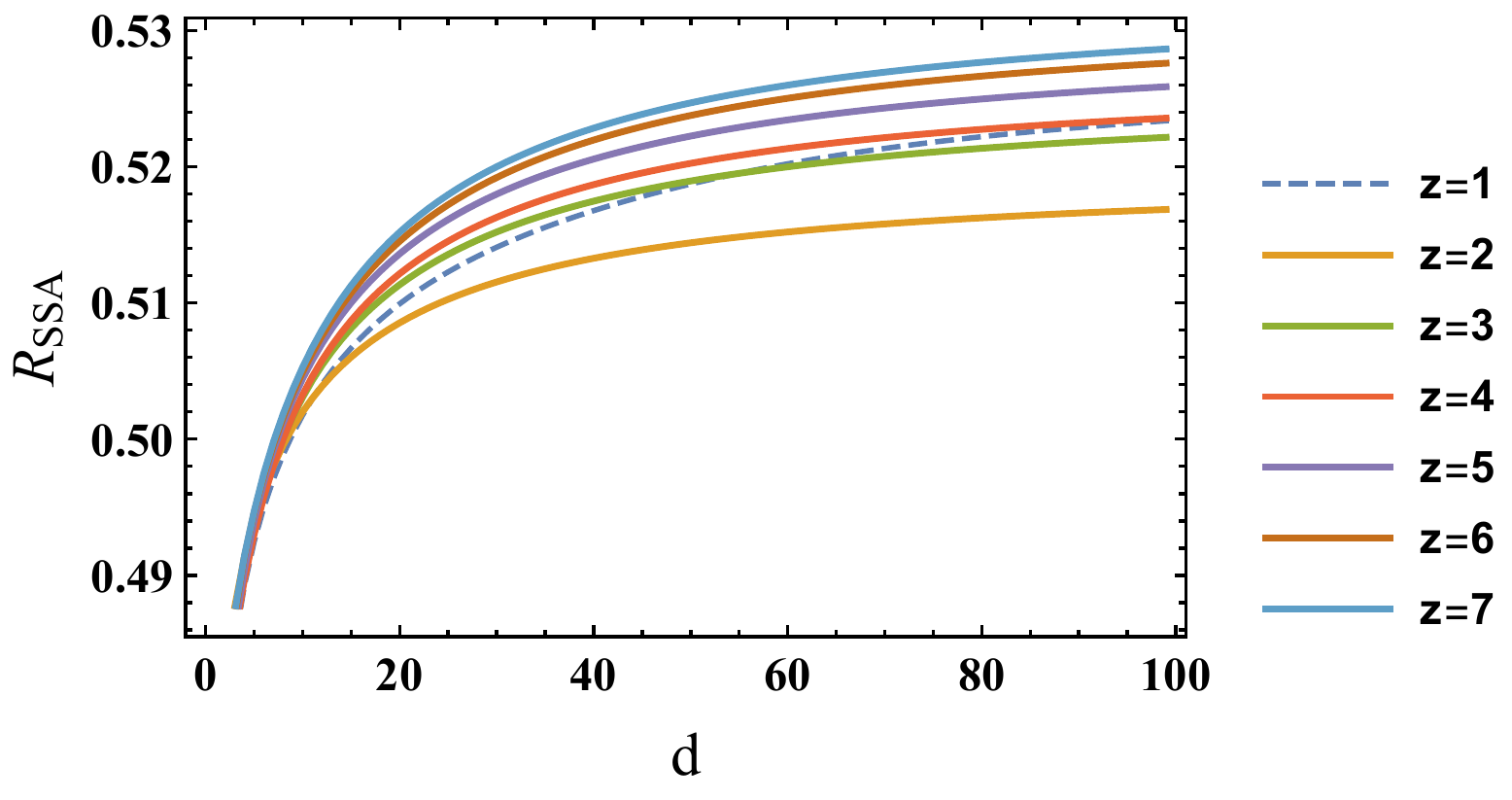}}
 \end{center}
 \caption{The breaking of strong superadditivity for $S_o$. Left: We plotted inverse temperature dependence of the cases with $L=2000, \ell_{A_1}=\ell_{A_2}=\ell_{B_1}=\ell_{B_2}=25, m=1.0\times 10^{-2}$ and the periodic boundary condition. Right: $z$-dependence of the same ratio as the left panels.}
 \label{fig:ssa}
\end{figure}
We would also like to check the strong superadditivity \eqref{eq:ssa}. To this end, we define
\be
R_{SSA}=\dfrac{S(A_1A_2:B_1B_2)}{S(A_1:B_1)+S(A_2:B_2)}. 
\ee
Figure \ref{fig:ssa} shows $S_o$ always violate the strong superadditivity. 
\section{Discussion}\label{sec:discussion}
As concluding remarks,  we would like to compare our results with holographic CFTs which was the original motivation to introduce the odd entropy.

We have seen how/when odd entropy can be larger than the usual von Neumann entropy by introducing thermal bath or Lifshitz-type generalization of the free theory. 
In particular, our results qualitatively suggest that large amounts of quantum correlations can make $S_o(A:B)-S(AB)$ positive. This is also qualitatively consistent with the holographic CFT. In there, we expect an inequality,
\be
S_o(A:B)-S(AB)\geq0,
\ee
always holds. We observed that this is also the case especially for the Lifshitz theories with even integers, where $S_o(A:B)-S(AB)$ and the logarithmic negativity matches. This is quite different behavior from the mutual information. Refer also to \cite{Angel-Ramelli:2020wfo} for analytical supports of similar setups in $2+1$ dimensions. Interestingly, in some of these cases, we can write down the vacuum wave functions explicitly. Therefore, it might be interesting to compare further these results with the ones for gravity theory where we have universal local correlations, so-called gravity edge modes\cite{Takayanagi:2019tvn}. Our result may also imply that the quantum corrections of holographic theories can be non-positive. 

However, we have to mention that once we introduce both thermal bath and Lifshitz parameters, we observed the above story becomes more complicated. It suggests that the above inequality highly constrains the theories to be ``holographic’’. To get a more quantitative understanding, we must further specify the role of quantum/classical correlations. We leave this issue for future work. 

We have obtained a series of numerical supports which ensures the odd entropy be a well-defined measure for mixed states. 
So far, our target was a very special class of field theories. Therefore, it would be fruitful to push forward this program further by studying more general many-body systems, like spin chains.

Previous to this work, similar aspects of the entanglement of purification has been studied in free scalar theories\cite{Bhattacharyya:2018sbw, Bhattacharyya:2019tsi}. Another interesting future direction is to study the reflected entropy as another holographic candidate for the entanglement wedge cross section\cite{reflected} as well as time evolution of these quantities\cite{te}. 

\section*{Acknowledgments}
We thank Cl\'ement Berthiere, Hugo A. Camargo, Jesse C. Cresswell, Michal P. Heller, Yuya Kusuki, Tadashi Takayanagi, and Koji Umemoto for useful discussion and interesting conversation. AM is generously supported by Alexander von Humboldt foundation via a postdoctoral fellowship. KT is supported by the Simons Foundation through the ``It from Qubit'' collaboration. KT is also supported by JSPS Grant-in-Aid for Research Activity start-up No.19K23441. AM would like to thank School of Particles and Accelerators at IPM where parts of this work was carried for their hospitality and access to their computer cluster. KT would like to thank Max-Planck-Institut for Physics and Gravitational Physics for the hospitality where the final stage of this work has been completed. We are also very grateful to workshop YITP-T-19-03 ``Quantum Information and String Theory 2019'' where this work was initiated.
\newpage
\appendix

\section{More on inflection points in odd entropy}\label{app:inf}
In this appendix, we study further on the inflection points of $S_o-S$ discussed in the main text. 
\subsection{A two qubit example}\label{app:twoq}

\begin{figure}[t]
 \begin{center}
 \resizebox{70mm}{!}{
 \includegraphics{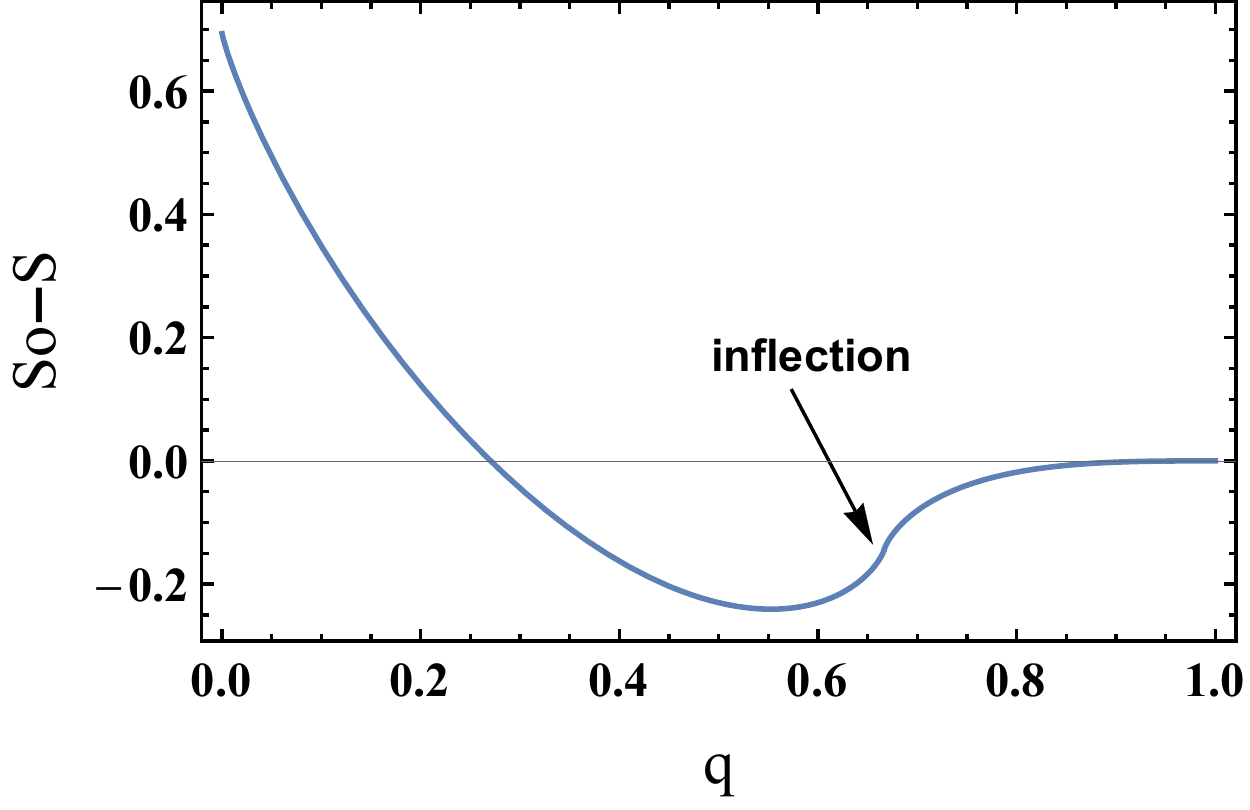}}
 \end{center}
 \caption{We have an inflection point at $q=2/3$. For $2/3<q\leq1$, we do not have any negative eigenvalues for $\rho_{AB}^{T_B}$. It means that we can say classical correlations dominate in this regime. }
 \label{fig:werner}
\end{figure}

Here we note one simple example of two-qubit systems from which we may learn the origin of the inflection point more easily. Let us consider a mixed state $\rho_{AB}$ acting on bi-partite Hilbert space $\mathcal{H}_A\otimes\mathcal{H}_B$,
\be
\rho_{AB}=(1-q)\ket{\Psi^-}\bra{\Psi^-}+\dfrac{q}{4}I_A\otimes I_B,
\ee
where $\ket{\Psi^-}$ is the EPR pair and $I_{A,B}$ are identity operator acting on $\mathcal{H}_{A,B}$.
For $\rho_{AB}^{T_B}$, we have the following eigenvalues,
\be
\mathrm{Spec}(\rho_{AB}^{T_B})=\left\{\dfrac{2-q}{4},\dfrac{2-q}{4},\dfrac{2-q}{4},\dfrac{3q-2}{4}\right\}.
\ee
From the PPT criterion\cite{Peres:1996dw} for two qubit systems\cite{Horodecki:1996nc}, this state is entangled for $0\leq q< 2/3$, whereas separable for $2/3\leq q \leq 1$. In Figure \ref{fig:werner}, we plotted the $S_o(A:B)-S(AB)$ for this example. 
\subsection{Thermal state with single interval}\label{app:thm1}

\begin{figure}[t]
 \begin{center}
  \resizebox{60mm}{!}{
 \includegraphics{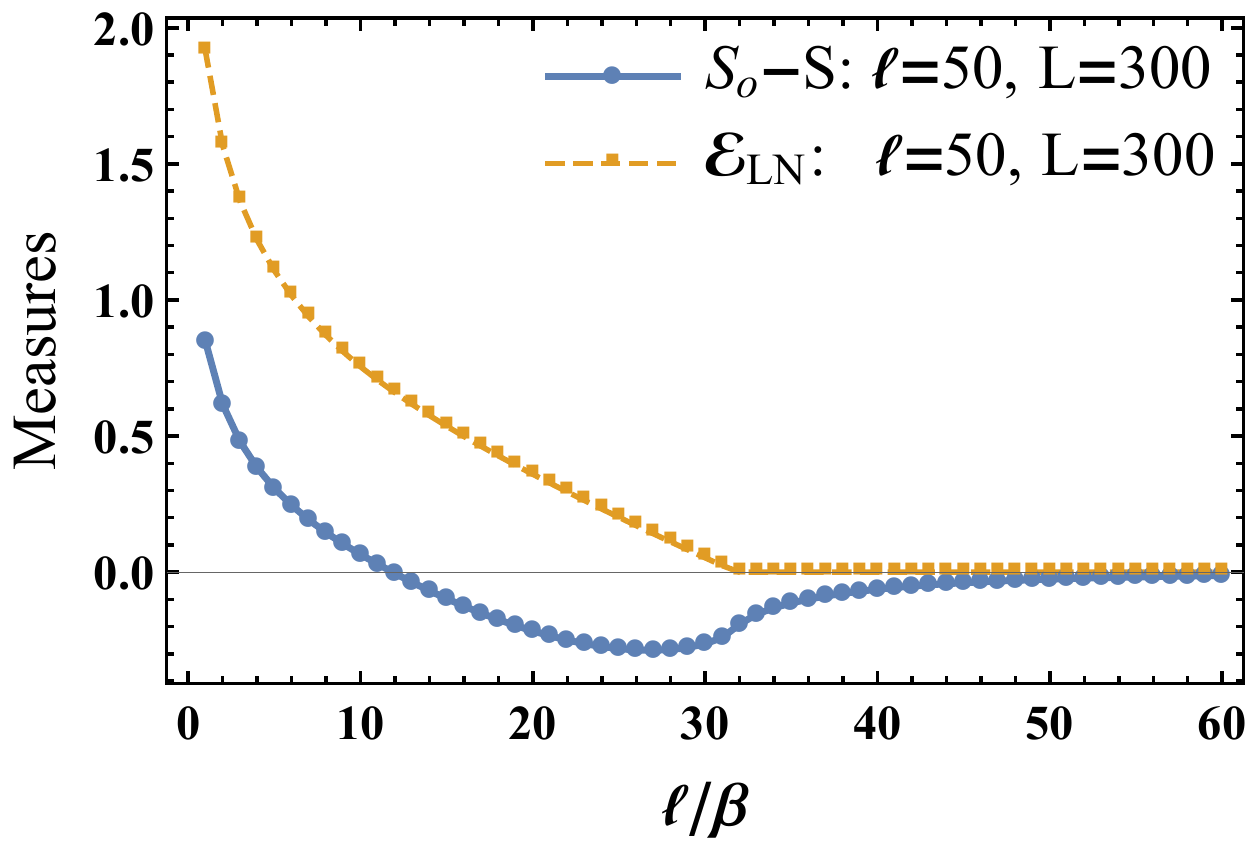}}
  \resizebox{62mm}{!}{
 \includegraphics{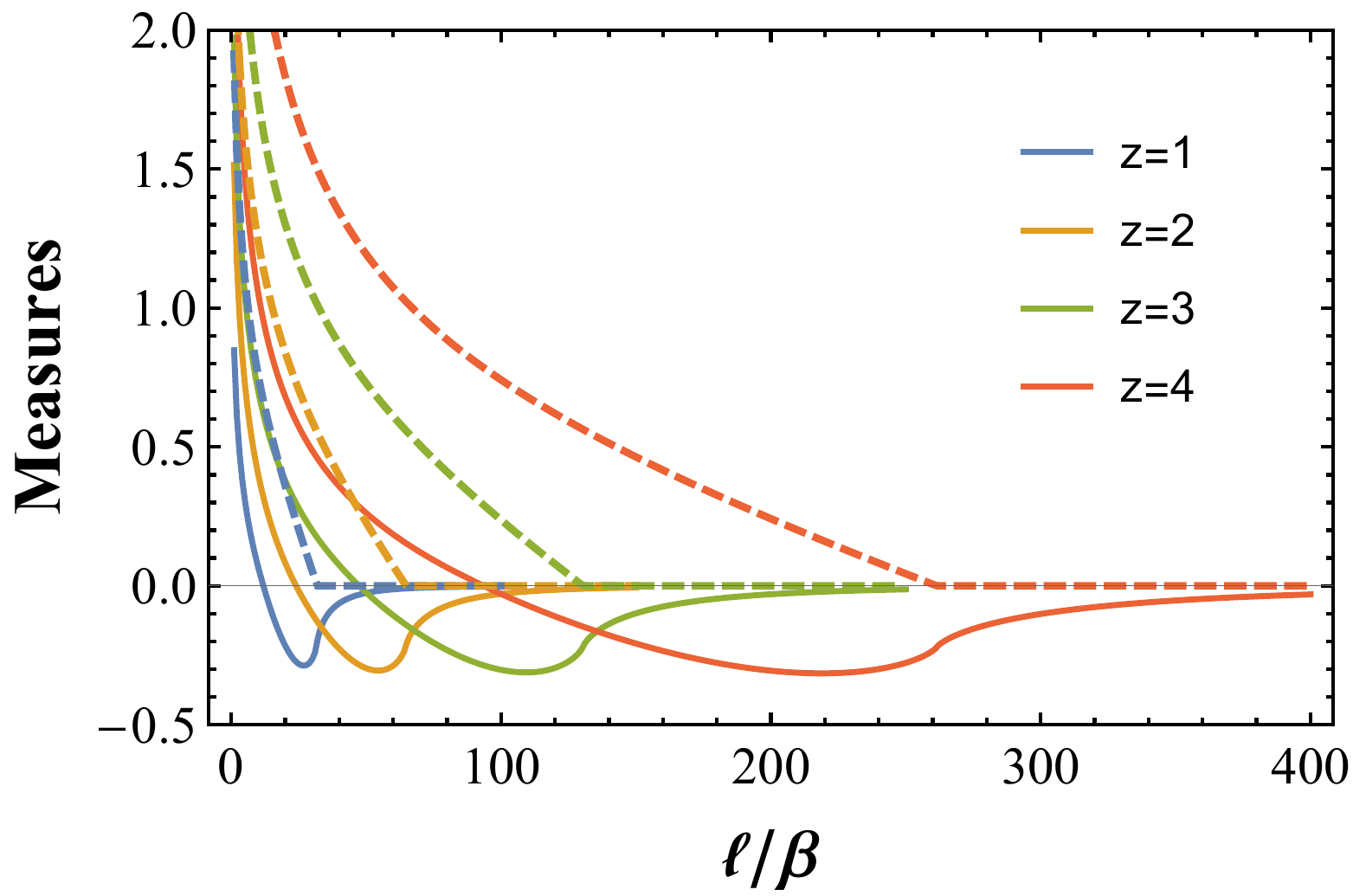}}
 \end{center}
 \caption{Left: $S_o-S$ and logarithmic negativity for thermal state with single interval. Here we fixed the subsystem size $\ell$ and plotted its temperature dependence. The $L$ corresponds to size of the total system. Right: The same setup with various value of $z$. The solid and dashed curves respectively correspond to $S_o-S$ and logarithmic negativity.}
 \label{fig:thm_single}
\end{figure}
\begin{figure}[t]
 \begin{center}
  \resizebox{140mm}{!}{
 \includegraphics{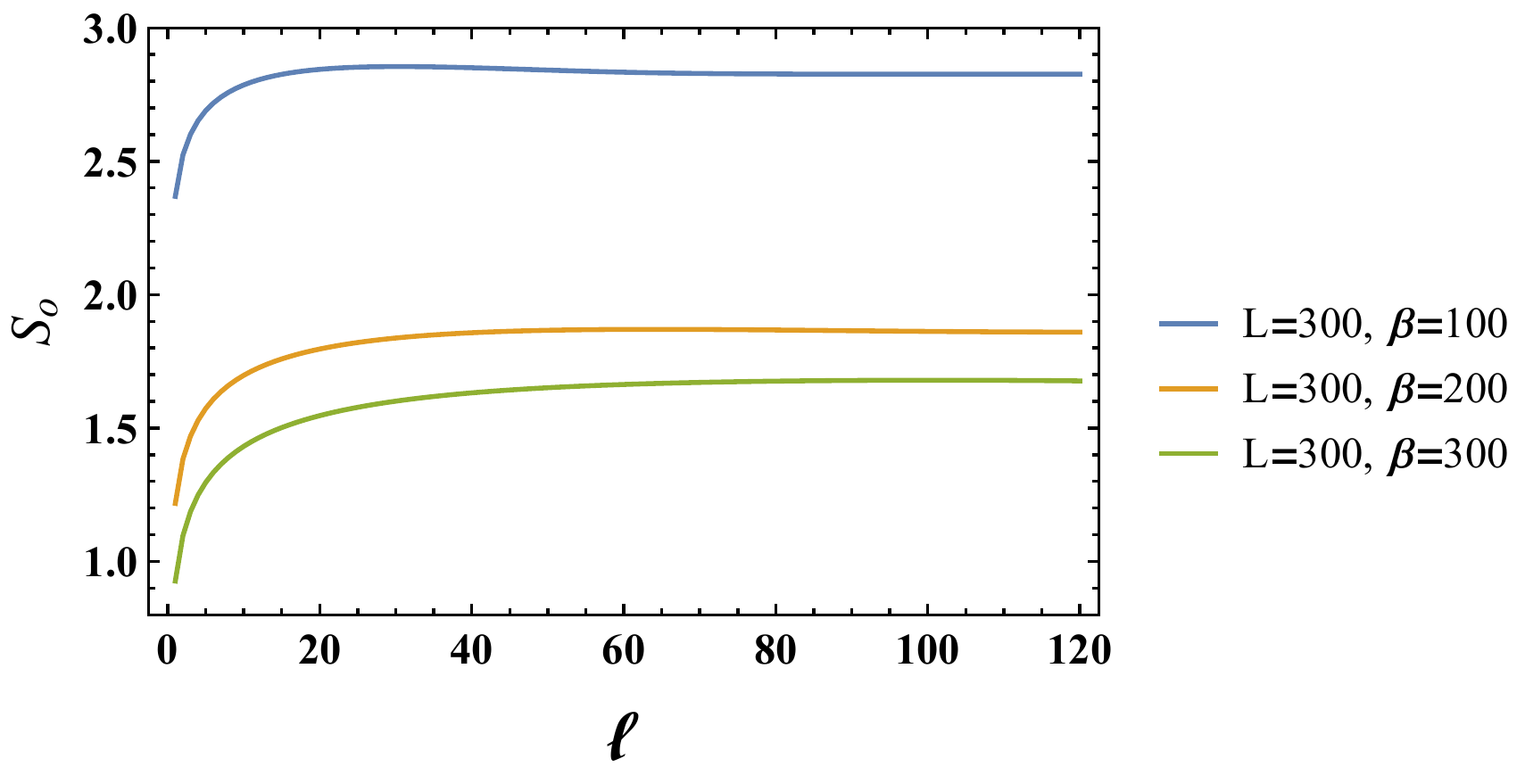}
 \includegraphics{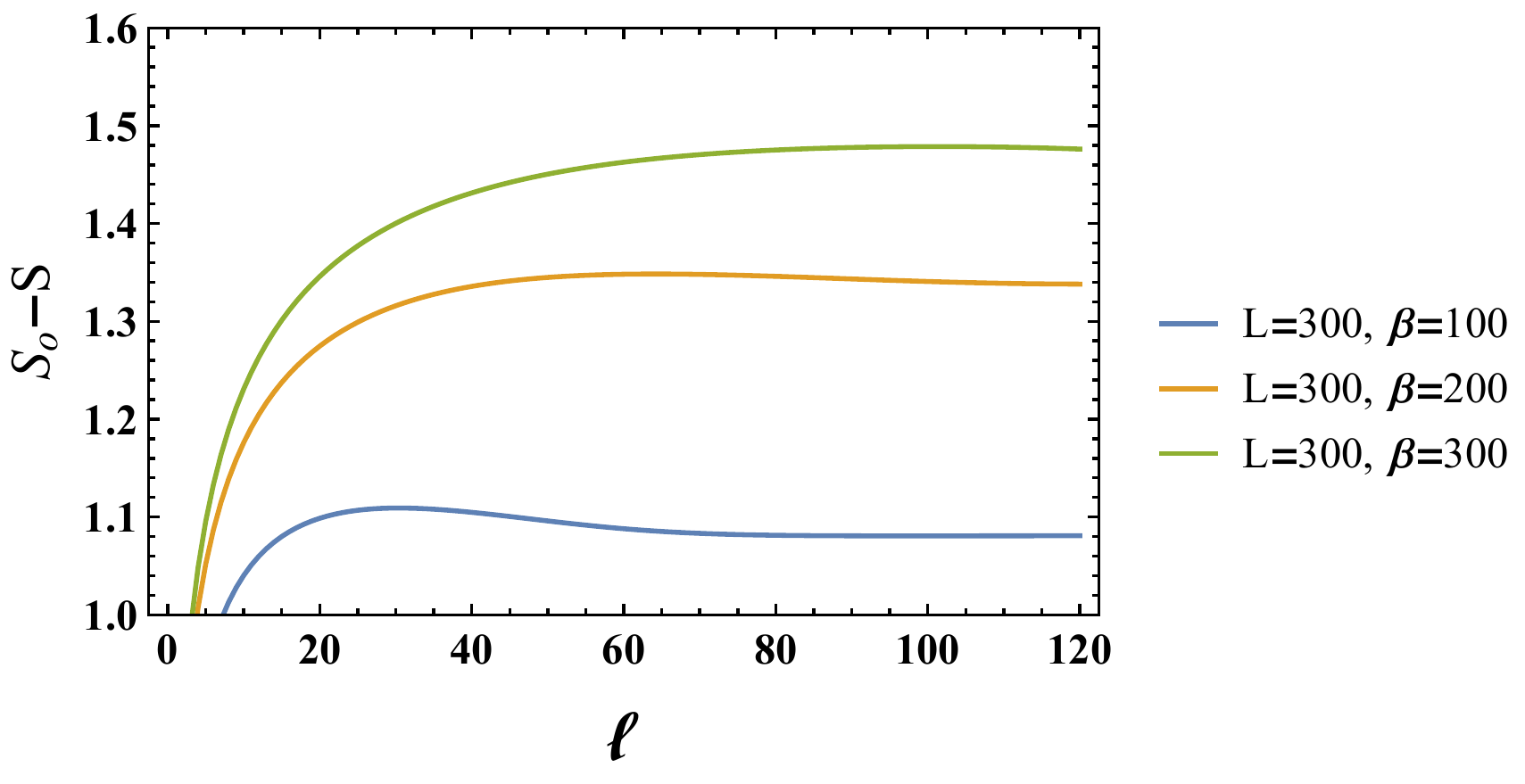}}
 \end{center}
 \caption{Plots for massless scalar field with Dirichlet boundary condition. The size of total system is given by $L=300$. The horizontal line labels the size of subsystem $\ell$. The hight of lump decreases as we decrease the temperature. Clearly, the origin of lumps are classical correlations. }
 \label{fig:thm_single2}
\end{figure}

We can see similar inflection points from our scalar field theories. Let us discuss the thermal state with a single interval case, namely $B=\bar{A}$. Note that, in this case, we cannot relate thermal results to vacuum ones via the naive conformal mapping. We plotted two figures where we fixed the subsystem size $\ell$ (Figure \ref{fig:thm_single}) or the inverse temperature $\beta$ (Figure \ref{fig:thm_single2}).

The shape of the curves in Figure \ref{fig:thm_single} is quite analogous to the previous qubit example. In particular, at the inflection point of $S_o-S$, the logarithmic negativity vanishes. It means that after this point, we have no negative eigenvalues for our partially transposed density matrix. In other words, the existence of such inflection point implies the decay of quantum correlations. The right panel where the deep of the minima of $S_o-S$ for different values of $z$ is almost unchanged, is nicely showing that increasing $z$ mainly increases quantum correlations. 

On the other hand, we can have another type of inflection point, after which the value of $S_o-S$ decreases. Let us see Figure \ref{fig:thm_single2}. For sufficiently large $\beta$ such that $\beta \gg a$, where $a$ is the lattice spacing and we took $a=1$, the $S_o$ increases monotonically under increasing the size of the subregion $\ell$. However, if the size of the subsystem approaches near but less than $L/2$, we observed a mild lump. After the lump, $S_o$ takes mostly a constant value which is reminiscent of a holographic result. Note that the von Neumann entropy $S(AB)$ takes constant value for every $\ell$. Interestingly, in this case, we observed that we obtain new negative eigenvalue contributions that do not effectively contribute before the present inflection point. Therefore, we may regard the inflection points in odd entropy as ``creation'' and ``annihilation'' of some quantum correlations. It might be an interesting future work to sharpen this observation so that we can learn more physical intuitions.

\begin{figure}[t]
 \begin{center}
  \resizebox{50mm}{!}{\includegraphics{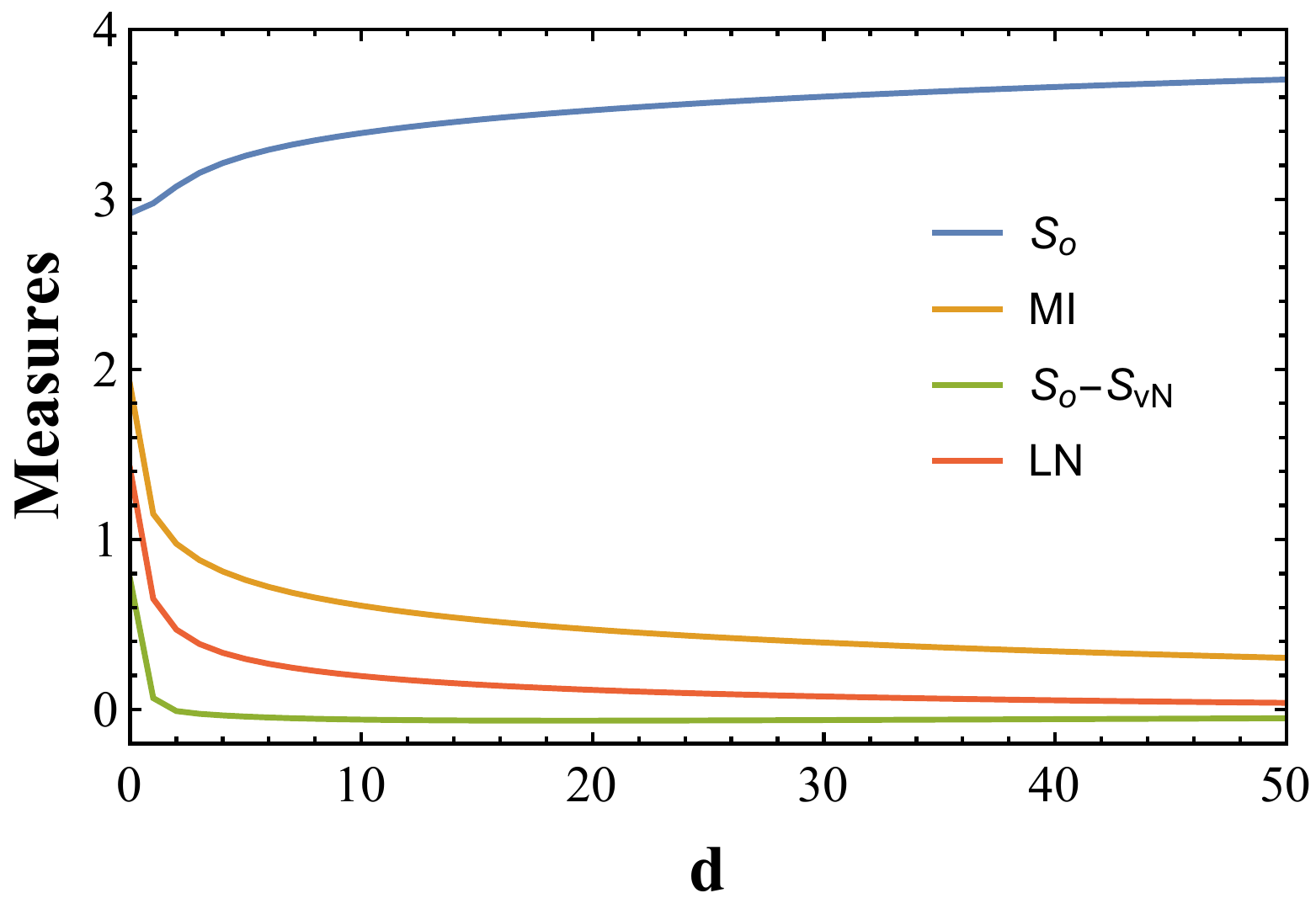}}
  \resizebox{50mm}{!}{\includegraphics{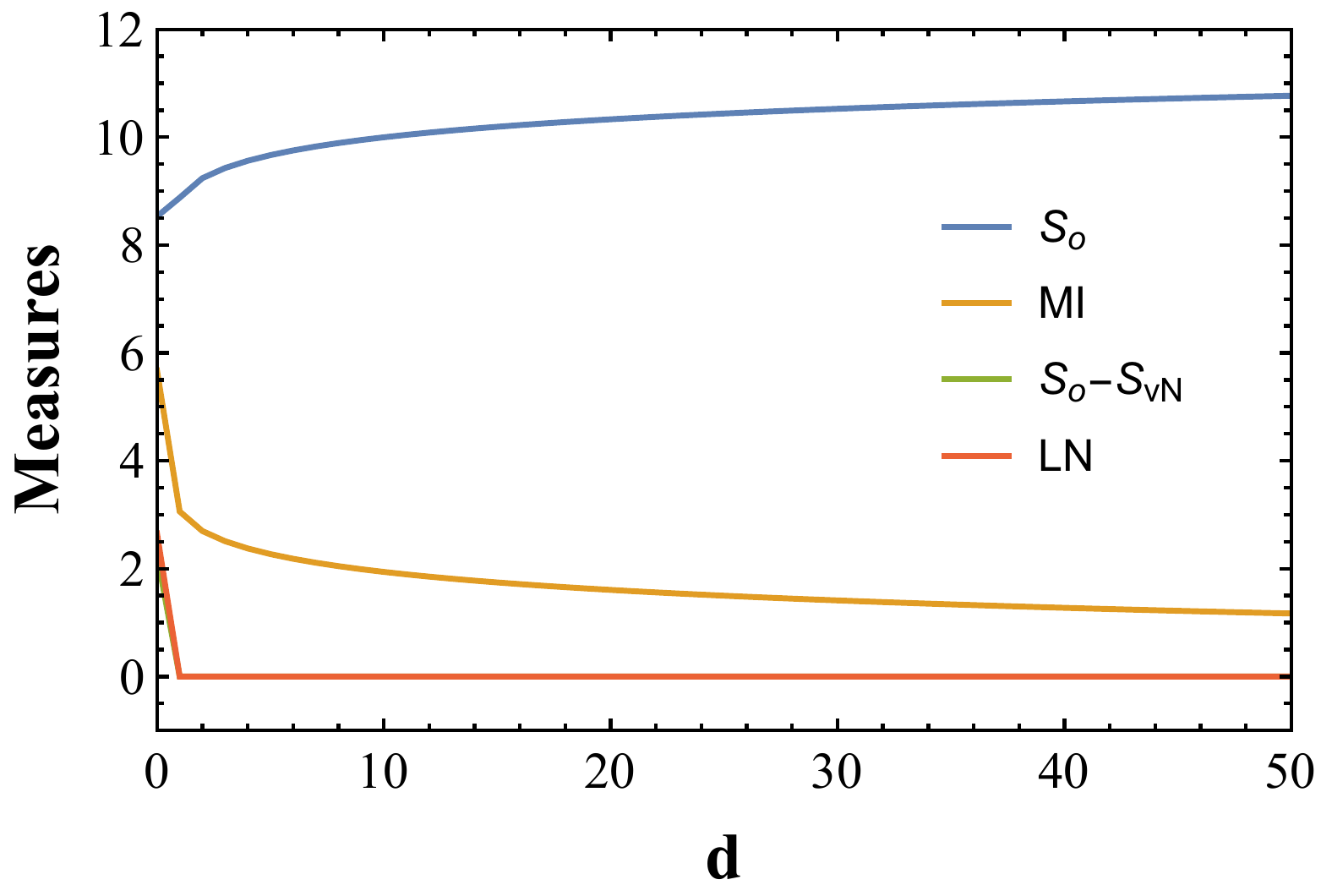}}
  \resizebox{50mm}{!}{\includegraphics{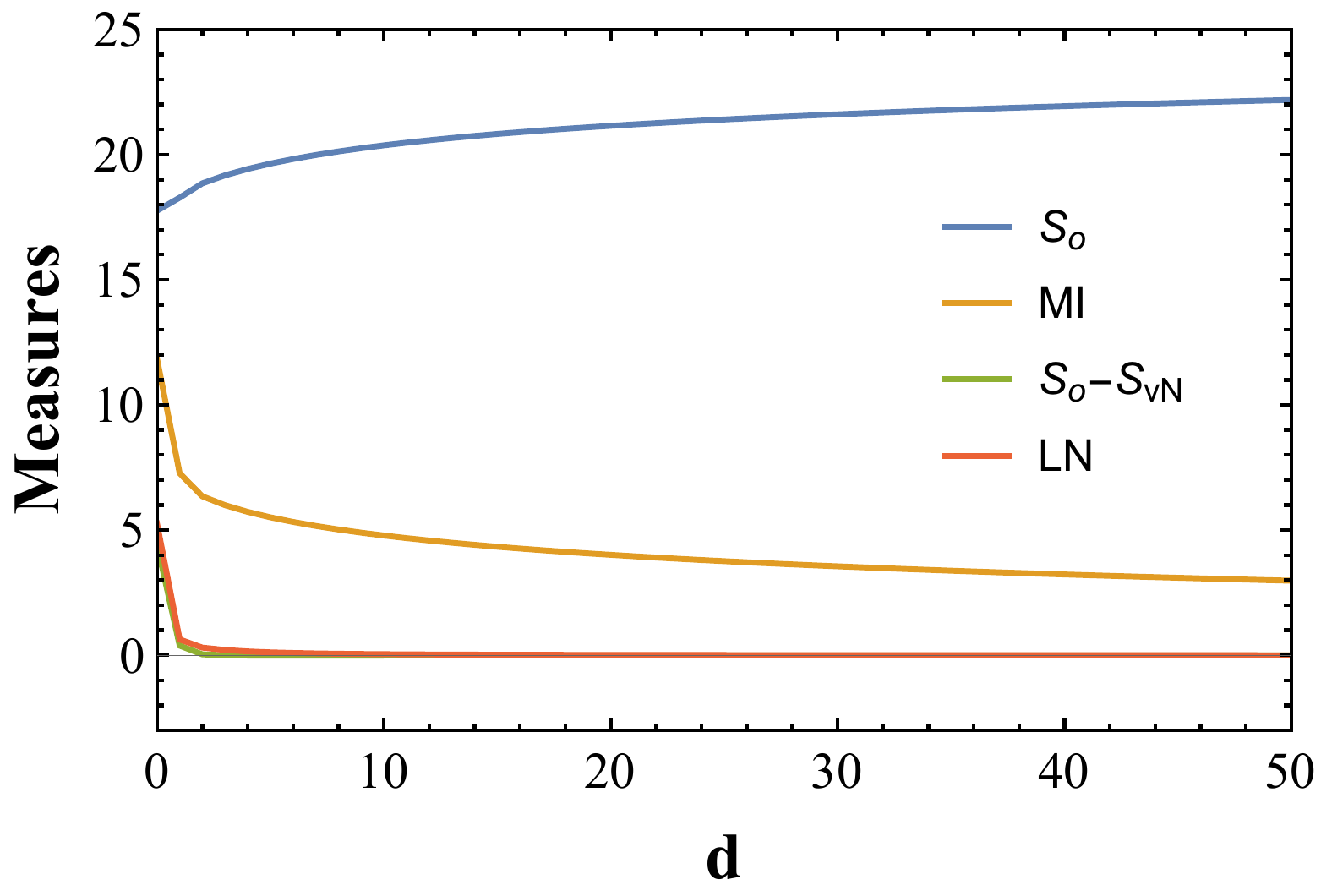}}
  \resizebox{50mm}{!}{\includegraphics{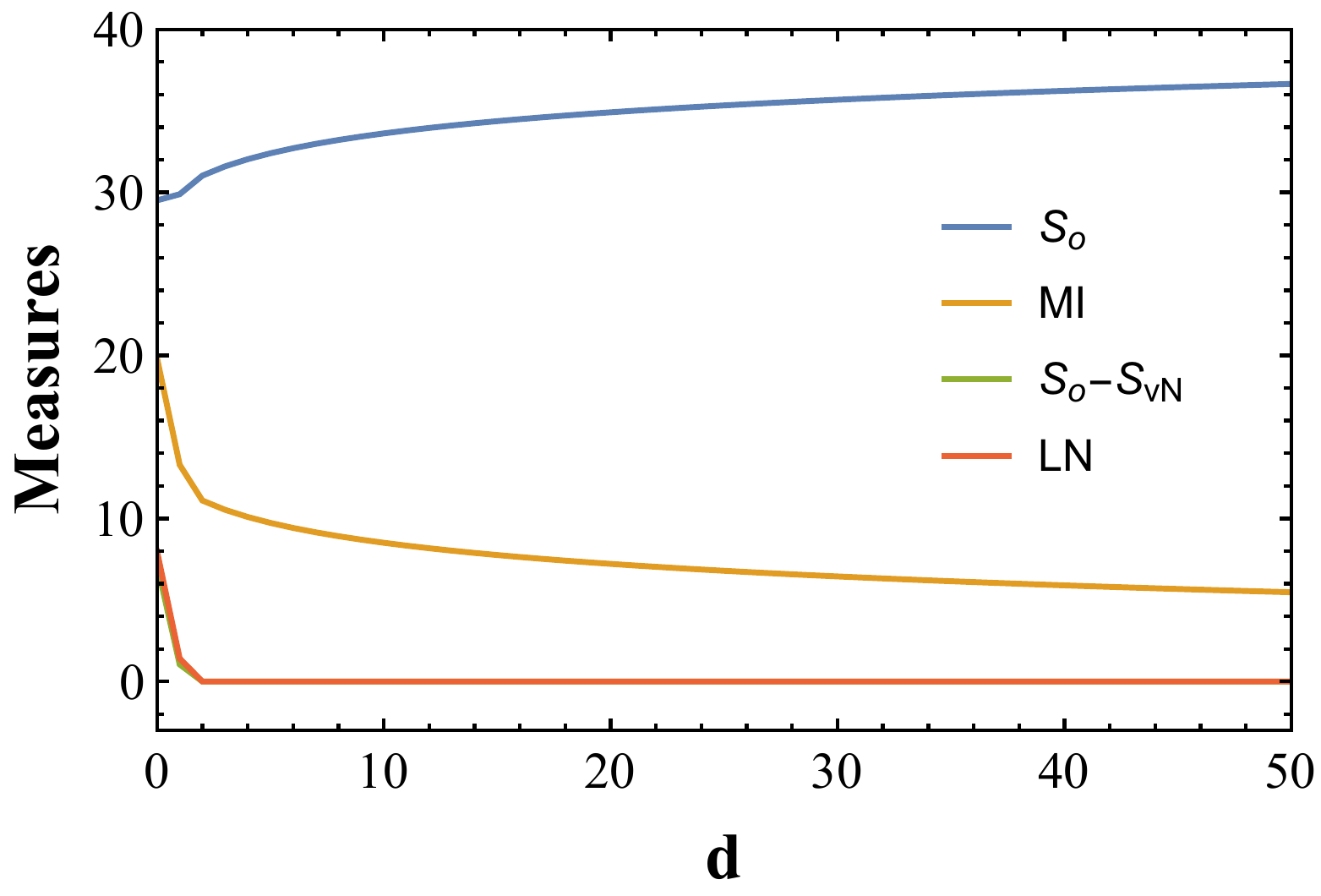}}
  \resizebox{50mm}{!}{\includegraphics{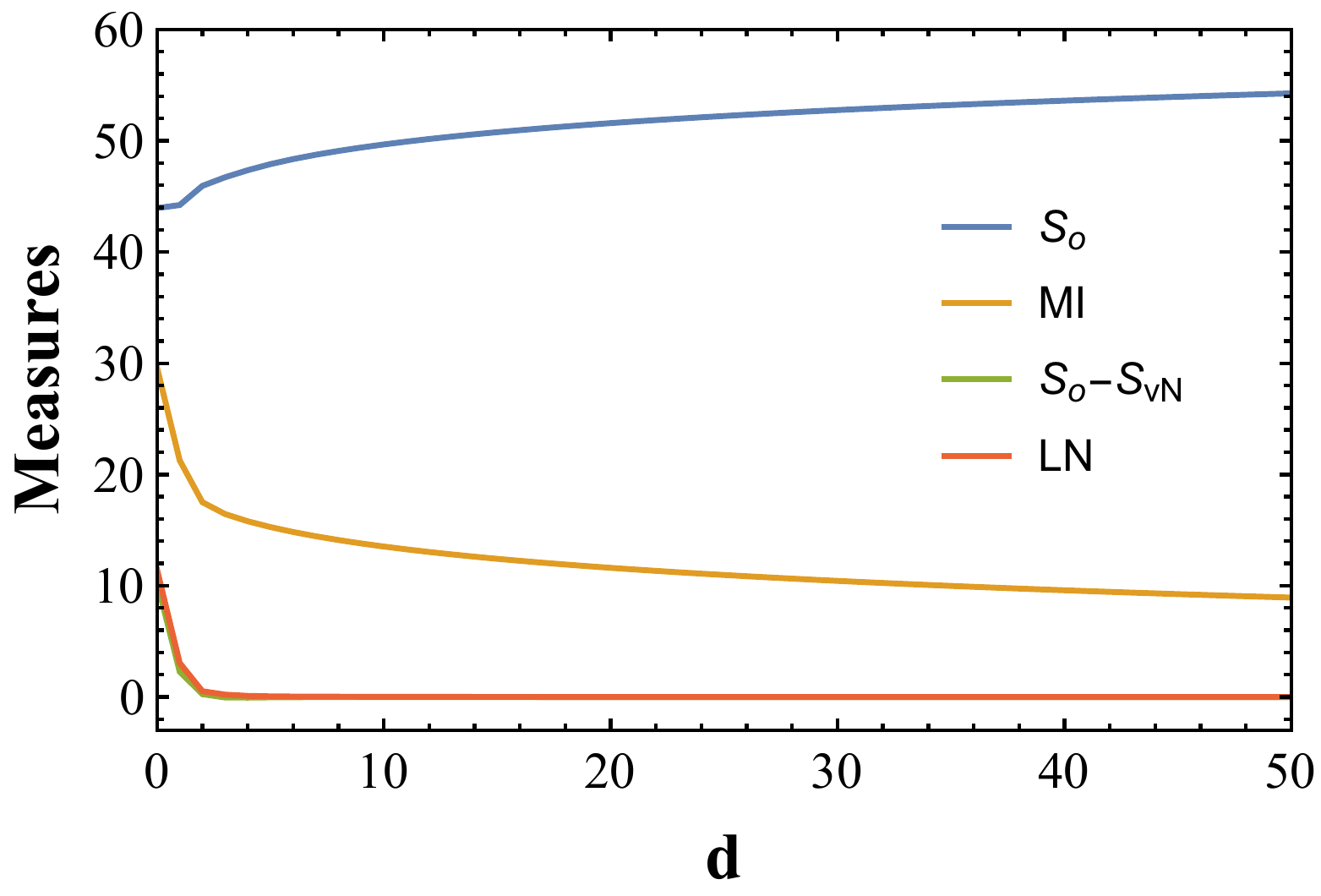}}
  \resizebox{50mm}{!}{\includegraphics{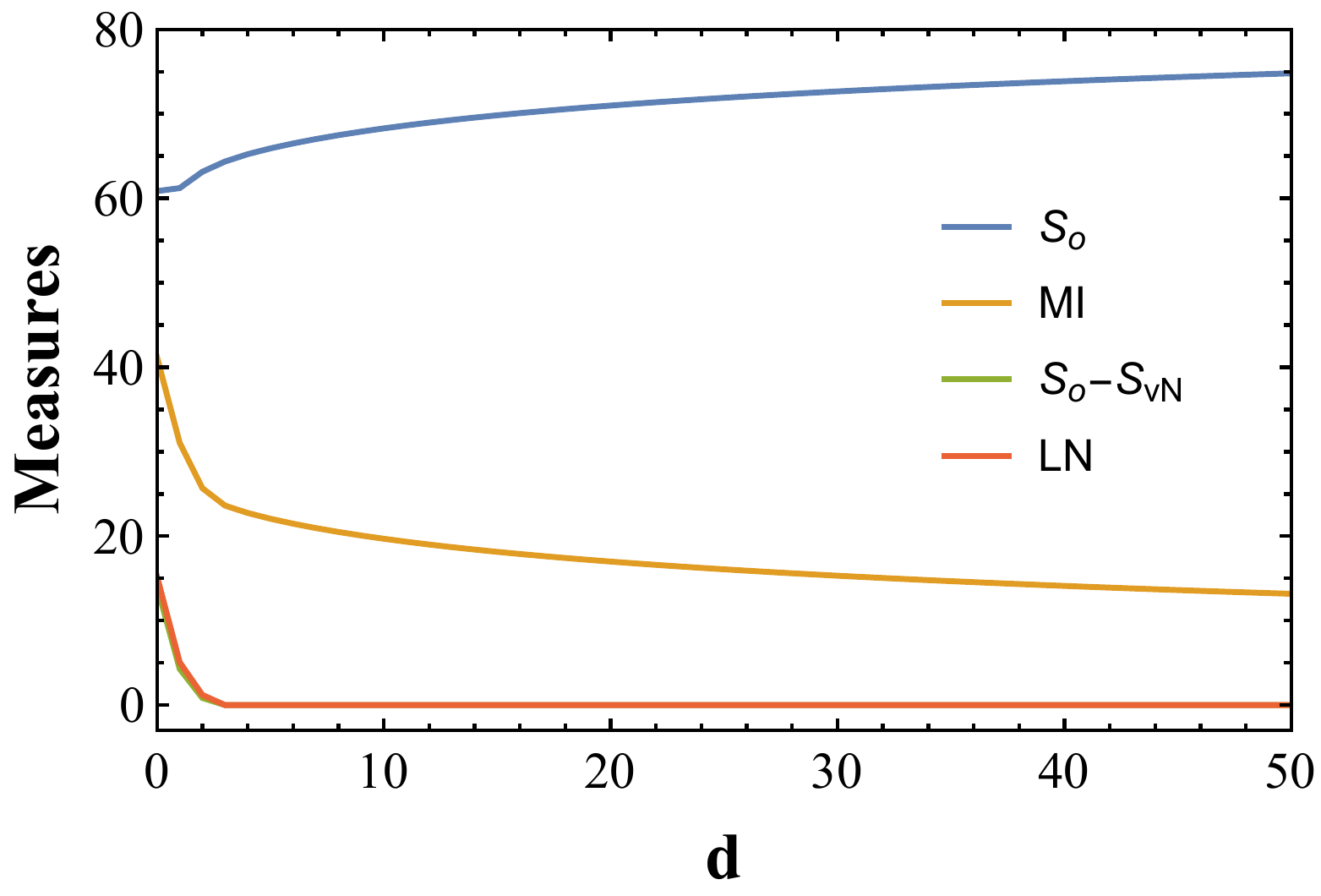}}
 \end{center}
 \caption{Comparison of measure for $z=1,2,3,4,5,6$ respectively starting from upper-left. In all plots we have set $\ell_1=\ell_2=100$ on a massless Dirichlet chain with $L=2000$.}
 \label{fig:measures}
\end{figure}

\section{More on comparison of different measures}\label{app:MI}

In this appendix, we show how the four measures we have studied, namely odd entropy $S_o$, deference between odd and von Neumann entropy $S_o-S$, mutual information $I$ and logarithmic negativity $\mathcal{E}_{LN}$ behave as functions of separation between $A$ and $B$. In Figure \ref{fig:measures}, we have presented all measures for different values of $z$. We can confirm the odd entropy is an entropy rather than a measure for total correlation between $A$ and $B$. It monotonically increases as a function of distance between subregions. As the value of dynamical exponent increases, the value of all these measures increases. 
As mentioned previously, the mutual information picks up more classical correlations than $\mathcal{E}_{LN}$ and $S_o-S$, especially in the large $d$ regime.

In the scale of odd entropy, there is a tiny difference between $S_o-S$ and $\mathcal{E}_{LN}$.  
For even values of $z$, these quantities coincide with each other. For odd values, as $z$ increases, due to the increase of quantum correlations, $S_o-S$ increases and for large enough values, it coincides with $\mathcal{E}_{LN}$.   


\end{document}